\documentclass[10pt,aps,prd,twocolumn,a4paper,showpacs,nofootinbib]{revtex4-1}
\usepackage{amsmath}
\usepackage{amsfonts}
\usepackage[colorlinks=true,linkcolor=blue,citecolor=blue,urlcolor=blue]{hyperref}
\usepackage{amssymb}
\usepackage{graphicx}
\usepackage{enumerate}
\usepackage{csquotes}

\providecommand{\abs}[1]{\left|#1\right|}
\providecommand{\ep}[1]{{e}^{#1}}

\newcommand{\phiin}{\phi_\omega^{\rm (in, co)}}
\newcommand{\phiina}{\phi_\omega^{\rm (in)}}
\newcommand{\phiinb}{\varphi_{-\omega}^{{\rm (in)}*}}
\newcommand{\phiout}{\phi_\omega^{\rm (out,co)}}
\newcommand{\phiouta}{\phi_\omega^{\rm (out)}}
\newcommand{\phioutb}{\varphi_{-\omega}^{{\rm (out)}*}}

\newcommand{\Sac}{A_\omega}
\newcommand{\Saa}{\alpha_\omega}
\newcommand{\Sab}{\beta_\omega}
\newcommand{\Sca}{\tilde{A}_\omega}
\newcommand{\Sbb}{\alpha_{-\omega}}
\newcommand{\Sba}{\beta_{-\omega}}
\newcommand{\Sbc}{B_\omega}
\newcommand{\Scb}{\tilde{B}_\omega}
\newcommand{\Scc}{\alpha_\omega^{\rm co}}
\newcommand{\na}{n_d}

\renewcommand{\Im}{{\rm Im}}
\renewcommand{\Re}{{\rm Re}}

\begin{document}

\title{Suppression of infrared instability in trans-sonic flows\texorpdfstring{\\}{}
by condensation of zero-frequency short wave length phonons}

\author{Xavier Busch}\email[]{xavier.busch@polytechnique.edu} 
\author{Florent Michel}\email[]{florent.michel@th.u-psud.fr} 
\author{Renaud Parentani}\email[]{renaud.parentani@th.u-psud.fr}
\affiliation{Laboratoire de Physique Th\'eorique, CNRS UMR 8627, B{\^{a}}timent\ 210,
 \\Universit\'e Paris-Sud 11, 91405 Orsay CEDEX, France}

\pacs{03.75.Kk, 04.62.+v, 04.70.Dy} 

\begin{abstract}
We analyze the peculiar infrared instability that characterizes stationary inhomogeneous flows when their velocity crosses the sound speed by decreasing values. For definiteness, we work in the context of one dimensional atomic Bose condensates. These flows are unstable under ultra low real frequency perturbations because of the unbounded mode amplification near the sonic horizon. This results in a condensation of low frequency phonons which produces a spatially structured flow in the supersonic domain. Numerical simulations reveal that this zero-frequency undulation suppresses the instability when its spatial extension is infinite, and when its phase is near that of a ``shadow soliton'' solution attached to the sonic horizon. These phenomena are akin to the condensation of rotons in flowing superfluid $^4$He when exceeding the Landau velocity. They also pertain to shallow water waves propagating on transcritical flows.
\end{abstract}

\maketitle

\section{Introduction}

The propagation of a fluid over an obstacle can produce many different patterns~\cite{Kamchatnov}, and this in several contexts such as longitudinal water flows in flumes~\cite{Johnson,Lawrence87,Wu87}, atomic Bose condensates~\cite{PhysRevA.68.063608,Hakim1997}, or polariton fluids~\cite{PhysRevLett.93.166401,50, Amo03062011}. In the present work, we focus on stationary trans-sonic flows which are supercritical in the upstream domain, and subcritical downstream. In these flows, the velocity $v$ crosses the speed of sound $c$ by decreasing values. When using the analogy with gravity~\cite{Unruh:1980cg,Unruh:1994je,Balbinot:2006ua,Barcelo:2005fc}, they correspond to white holes since counter-propagating waves are blueshifted when approaching the sonic horizon $v = c$~\cite{Macher:2009tw,Macher:2009nz}. Because of this blueshift, the flows display an intriguing infrared instability which results in the formation of a macroscopic zero-frequency wave with a short wave length, see Refs.~\cite{Mayoral:2010ck,Coutant:2012mf}. In this paper we first complete these works, and then show that the formation of this wave suppresses the instability when its phase is close to that of a shadow soliton solution attached to the sonic horizon. For definiteness we work with one dimensional atomic Bose condensates but our results apply to the aforementioned other fluids up to minor modifications. 

To start, we study the space of solutions of the Gross-Pitaevskii equation (GPE) which describe stationary trans-sonic flows. When restricting attention to asymptotically homogeneous flows in the downstream region, regular solutions of the GPE are characterized by the amplitude of the density modulation in the upstream region. In analogy with what is found in water flumes, we shall call this stationary structure an \textit{undulation}. To linear order, for a local constriction, the general solution can be conceived as a superposition of a \enquote{free} solution and a \enquote{forced} one. When considering fully non-linear solutions of the GPE, this superposition loses its meaning, but the space of solutions is still characterized by the asymptotic amplitude of the undulation.

We then study the scattering of phonon modes, solutions of the Bogoliubov-de~Gennes equation, in stationary trans-sonic flows characterized by the amplitudes of the free and the forced undulations. These flows are energetically unstable since they possess negative energy excitations. In appendix~\ref{app:dynstab}, we verify that they are dynamically stable, as the spectrum of asymptotically bounded modes contains no complex frequency~\cite{Macher:2009nz,Mayoral:2010ck}, at least when the amplitude and the extension of the undulation are small enough. When there is no undulation, the mode amplification associated with the scattering of real frequency modes diverges in the zero frequency limit. As a result, random ultra low frequency perturbations give rise to an undulation~\cite{Mayoral:2010ck,Coutant:2011in,Coutant:2012mf,Coutant:2012zh}. When the dispersion is of the Bogoliubov type, the undulation propagates from the sonic horizon in the supersonic region, whereas it propagates in the subsonic region when the dispersion is \enquote{normal}. Interestingly, this infrared (IR) instability is akin to that described by Pitaevskii~\cite{Pitaevskii1984} which occurs in homogeneous flows of superfluid $^4$He when their velocity exceeds the Landau velocity, and which is stabilized by a condensation of zero frequency rotons. In the present case, the instability results in the condensation of short wave length ultra low frequency phonons. As in~\cite{Pitaevskii1984}, this produces a structured flow characterized by the wave number which is the non trivial zero-frequency solution of the dispersion relation. In the present case, the structure precisely corresponds to a {\it free} undulation. In a linearized treatment, its sign is undetermined. The degeneracy is lifted when including non-linear effects: the stable solution corresponds to a shadow soliton solution, and the unstable one to a soliton solution. This is in agreement with what has been found when studying the black hole laser instability~\cite{Michel:2013wpa} in flows with two sonic horizons.

In the second part of this work, using numerical techniques adapted from those of~\cite{Macher:2009nz,Finazzi:2010yq,Finazzi:2012iu}, we show that short undulations reduce the IR instability when their phase is close to that of the shadow soliton, and amplify it when it is close to that of the soliton solution. In addition, when considering infinitely long undulations, the mode amplification is eventually suppressed for the shadow soliton phase, whereas the soliton phase gives rise to a dynamical instability.  These results should also apply to water waves propagating on top of trans-critical flows in flumes. It should thus be possible to test them in experiments similar to the recent ones aimed at detecting the analogue Hawking radiation in atomic BEC~\cite{Lahav:2009wx}, and in flumes~\cite{Weinfurtner:2010nu,Rousseaux:2007is}. It should be also pointed out that this stabilization differs from that found in $^4$He. Indeed, the energetic instability is maintained, since the spectrum still contains negative energy excitations, but the mode amplification in the zero-frequency limit no longer diverges.

This paper is organized as follows. In Sec.~\ref{sec:NLsol}, we analyze inhomogeneous solutions of the GPE which describe trans-sonic flows. We also add a localized obstacle and briefly discuss the space of solutions. In Sec.~\ref{sec:lin}, we numerically solve the Bogoliubov-de~Gennes equation in these backgrounds and study the infrared instability, and its growth or suppression due to the undulation. We conclude in Sec.~\ref{sec:concl}. Appendix~\ref{App:homo} recalls the basic properties of homogeneous condensates. In appendix~\ref{app:dynstab}, we establish that flows with small undulations are dynamically stable, and relate the appearance of a dynamical instability to the end point evolution to the growth of the IR instability. In appendix~\ref{App:smallobs} we analytically compute the forced undulation due to a small obstacle. 

\section{Sonic analogue white hole flows and undulations} 
\label{sec:NLsol}

\subsection{Inhomogeneous Gross-Pitaevskii equation}

In this subsection we study trans-sonic stationary flows, solutions of the GPE for a particular class of potentials. The basic properties of flows in homogeneous potentials are reminded in appendix~\ref{App:homo}.

Because we are interested in stationary solutions, it is appropriate to work with the mean density $\rho(z)$ of condensed atoms and their velocity $v(z)$. The one dimensional GPE then gives 
\begin{equation}
\begin{split}
\rho v &= J , \\
\frac{\hbar^2 }{2 m} \frac{\partial_z^2\sqrt{\rho}}{\sqrt{\rho}} &= V + g \rho + \frac{m J^2}{2 \rho^2},
\label{GPE1}
\end{split}
\end{equation}
where $J$, taken positive, gives the constant value of the flux. Here $V(z)$ is the external potential, and $g$ the $z$-independent two-body interaction constant. To simplify the expressions, we work in a unit system where $\hbar = 2 m = 2 g J = 1$. In these units, the healing length $\xi$ is 
\begin{equation}
\begin{split}
\label{rhoc}
\xi(z) &= \left( \frac{\rho(z)}{\rho_c} \right)^{-1/2}, 
\end{split}
\end{equation}
where $\rho_c \equiv J $. Similarly, the adimensional flow velocity and sound speed $c^2 = g \rho/ m$ are given by
\begin{equation}
\begin{split}
\label{eq:vandc} 
v &= \frac{\rho_c}{\rho}, \quad c = \sqrt{\frac{\rho}{\rho_c}} .
\end{split}
\end{equation}
Then the ratio $F = v/c$ reads~\footnote{Notice the similarity with the case of surface gravity waves in hydrodynamics where this ratio (called Froude number) is given by $F = h^{-3/2}$, $h$ being the adimensional water height~\cite{Michel:2014zsa}. In fact most of the forthcoming analysis applies to surface waves under the substitution $\rho \to h$, and the replacement of Eq.~\eqref{GPE2} by the Korteweg-de~Vries equation~\cite{Johnson}.} 
\begin{equation}
\begin{split}
F(z) = \frac{v(z)}{c(z)}= \left( \frac{\rho_c}{\rho(z)} \right)^{3/2}. 
\end{split}
\end{equation}

In this system, Eq.~\eqref{GPE1} gives 
\begin{equation} 
\begin{split}
2 \rho \rho'' - 4 V \rho^2-\rho'^2-2 \frac{\rho^3}{\rho_c}-\rho_c^2=0 , 
\label{GPE2}
\end{split}
\end{equation}
where a prime denotes a derivative with respect to $z$. For reasons of simplicity, we now consider the set of reference profiles given by
\begin{equation}
\begin{split}\label{eq:V}
V_0(z) = \frac{2 \rho_0 \rho_0'' - \rho_c^2-2 {\rho_0^3}/{\rho_c}-(\rho_0')^2}{4 \rho_0^2}, 
\end{split}
\end{equation}
where 
\begin{equation}
\label{eq:rhotanh}
\begin{split}
\rho_{0}(z) = \bar \rho \left(d \tanh \left(\frac{2 \kappa z}{3 d}\right)+1\right).
\end{split}
\end{equation}
By construction, $\rho_0$ is an exact solution of Eq.~\eqref{GPE2}. It is characterized by the central value of the density $\bar \rho$, the spatial gradient $\kappa = (3/2) \partial_z \ln \rho $ evaluated at $z= 0$, and the asymptotic values $\rho_p \equiv \bar \rho (1 - d)$ in the supersonic region, and $\rho_b \equiv \bar \rho (1 + d)$ in the subsonic one. To further reduce the number of free parameters, in numerical analysis, we only consider the case where $\bar{\rho} = \rho_c$, so that the sonic horizon $v/c= 1$ is located at $z = 0$. These simplifications do not restrict the validity of our results. The essential point is that $F$ crosses $1$ by decreasing values in the direction of the flow. 

When $\bar{\rho} = \rho_c $, the analogue surface gravity~\cite{Unruh:1980cg,Barcelo:2005fc} (i.e. the frequency $\kappa_G \equiv \partial_z (v - c)$ evaluated at the sonic horizon) is given by $ \kappa_G = - \kappa $ when $J>0$. Its negative value confirms that we are dealing with an analogue white hole flow, i.e., a trans-sonic one which becomes subsonic along the direction of the flow. It corresponds to a white hole (and not a black hole) geometry because low wave number phonons propagating against the flow are blueshifted when approaching the sonic horizon at $z = 0$. The surface gravity $\kappa_G$ is highly relevant in that it governs the scattering of low frequency phonons in smooth trans-sonic flows~\cite{Unruh:1980cg,Unruh:1994je,Brout:1995wp}. More precisely, when there is a neat scale separation between the surface gravity and the healing length, i.e. when $\kappa \xi/c \ll 1$, the spectrum of phonons spontaneously emitted by the scattering on the sonic horizon is, to a very good approximation~\cite{Macher:2009nz,Finazzi:2010yq,Finazzi:2012iu}, a Planck spectrum with a temperature given by $T_H = \kappa/2\pi$, in accordance with the famous prediction of Hawking~\cite{Hawking:1974sw}.

Before studying this scattering and its implications, we classify the stationary solutions of the GPE which are smoothly connected to our reference solution of Eq.~\eqref{eq:rhotanh}.

\subsection{Free and forced density perturbations}

\begin{figure*}
\begin{minipage}{0.47\linewidth}
\includegraphics[width=1\linewidth]{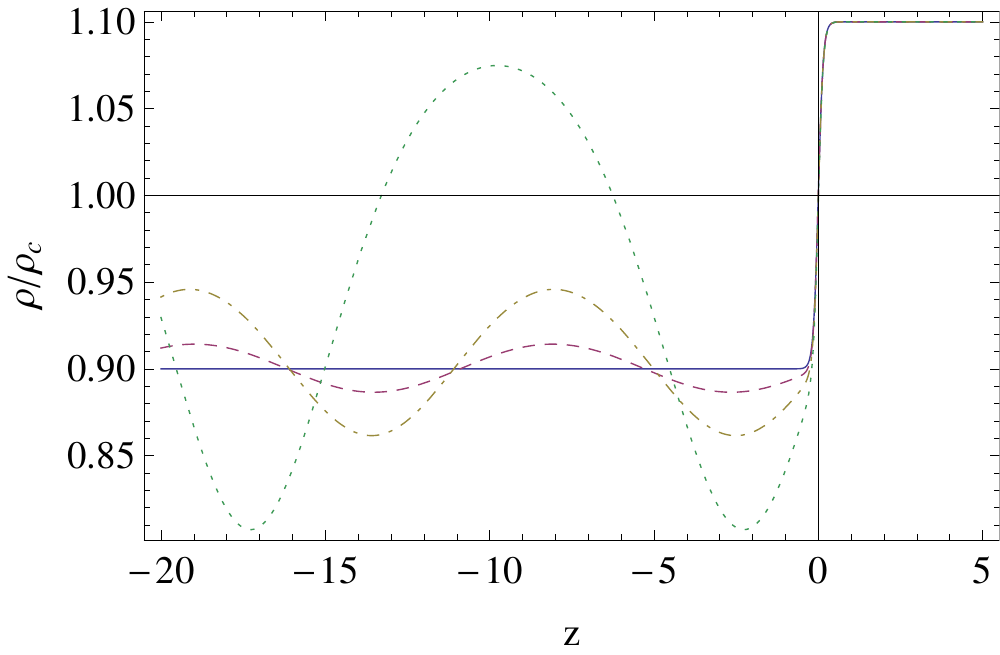}
\end{minipage}
\hspace{0.03\linewidth}
\begin{minipage}{0.47\linewidth}
\includegraphics[width= \linewidth]{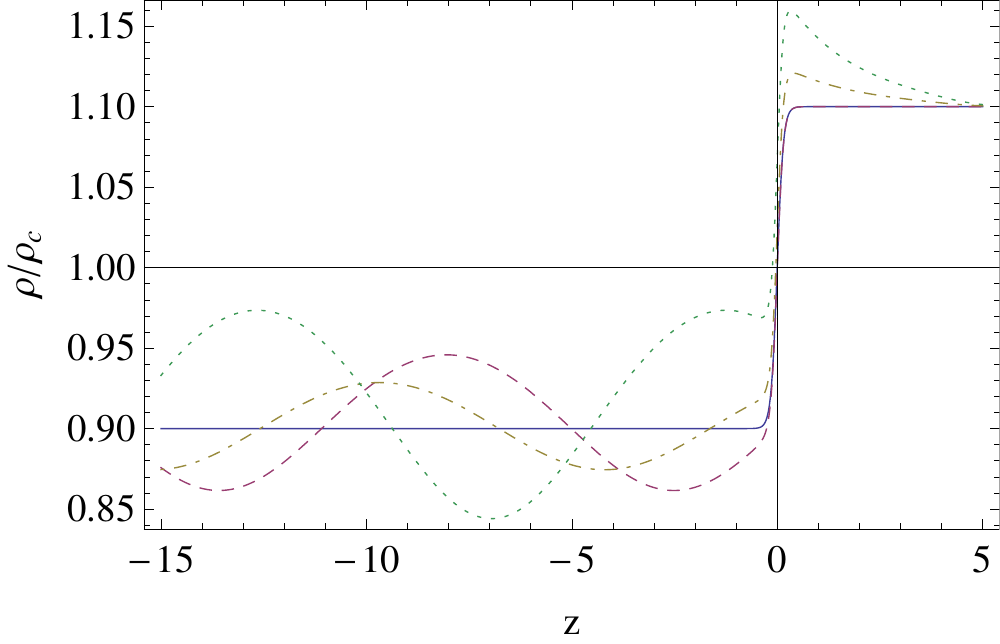}
\end{minipage}
\caption{On the left panel, in units of $\rho_c$ of Eq.~\eqref{rhoc}, we represent the density profile of Eq.~\eqref{eq:rhotanh} (blue solid) with $\kappa = 10 d = 1$, and those of three forced undulations, as functions of the distance in units of the healing length evaluated at $z= 0$. The background flow is supersonic for $z < 0$, and the sonic horizon is located at $z = 0$, where $\rho_0/\rho_c = 1$. The amplitudes of the obstacle of Eq.~\eqref{dV} are $\bar \delta_V = 0.01$ (purple dashed), $0.03$ (yellow dotted dashed) and $0.08$ (green dotted), while the other parameters are $z_p = 0$, $\kappa_p = 5$. We notice that the wavelength of the oscillations increases when the undulation amplitude leaves the linear regime, here for $\bar \delta_V \gtrsim 0.05$. For larger amplitudes, the undulation produces pairs of sonic horizons when $\rho_0/\rho_c = 1$, see the dotted curve. On the right panel, we represent $\rho(z)/\rho_c$ for an obstacle with $\bar \delta_V = 0.03$, with the forced undulation (dashed) and when free undulations with different amplitudes are added on top of it: $A_h = 0.035$ (dash-dotted) and $A_h = 0.088$ (dotted). One clearly sees that $\varphi_r$, the phase of the resulting undulation, differs from that of the forced one.}
\label{fig:rhoforvarV}
\end{figure*}
To characterize this set of solutions, we consider perturbed potentials given by $V= V_0(z) + \delta V(z)$. To describe local obstacles, we impose that $\delta V$ vanishes in both asymptotic regions. When using analytical methods, we shall study the case where $\delta V \ll V_0$, so that a linear expansion in $\delta V/V_0$ can be made. However, when using numerical methods, this inequality can be relaxed without difficulty, and the non-linear aspects of the GPE can be fully taken into account. 

We write $\rho = \rho_0 (1 +\delta \rho)$, with $\rho_0(z)$ given in Eq.~\eqref{eq:rhotanh}. To linear order, the relative deviation $\delta \rho$ obeys 
\begin{equation}
\label{eq:forcelineqrho}
\begin{split}
\delta \rho ''+\frac{ \rho_0 '}{\rho_0} \delta \rho ' - \delta \rho \left(\frac{\rho_0}{\rho_c} - \frac{\rho_c^2}{\rho_0^2}\right) = 2 \delta V. 
\end{split}
\end{equation}
The general solution is the sum of a {\it free} perturbation $\delta \rho_{\rm hom}$, solution of the homogeneous equation, and a particular solution of Eq.~\eqref{eq:forcelineqrho}, $\delta \rho_{\rm forced}$. To fix without ambiguity this {\it {forced}} solution, it is mathematically convenient to impose that it vanishes on the right of the perturbation, in the downstream subsonic region. (This is also physically justified as low frequency perturbations with large wave-vectors propagate upstream in the supersonic region, as explained in appendix~\ref{App:homo}.) On the asymptotic left supersonic region, this forced solution is oscillatory 
\begin{equation}
\label{eq:deltarhoforced}
\begin{split}
\delta \rho_{\rm forced} \sim A_f \sin [ k_0 z + \varphi_f] ,
\end{split}
\end{equation}
where the asymptotic wave number $k_0$ is 
\begin{equation}
\label{eq:defku}
\begin{split}
k_0 = \frac{\sqrt{ \rho_c^3-\rho_p^3 }}{\rho_p \sqrt{\rho_c}} = \sqrt{v(z \rightarrow -\infty)^2 - c(z \rightarrow -\infty)^2}. 
\end{split}
\end{equation}
Indeed, in both asymptotic regions, the solutions of Eq.~\eqref{eq:forcelineqrho} are superpositions of exponential functions, with wave-vectors $k$ solutions of the dispersion relation Eq.~\eqref{eq:disprel} for $\omega = 0$. In the supersonic region, the modes oscillate since $k_0$ is real. Instead, in the right subsonic region, one gets exponentially decaying and growing functions $e^{\pm \lambda_0 z}$, with $\lambda_0 = \sqrt{c^2 - v^2}$ for $z \to \infty$.

In Fig.~\ref{fig:rhoforvarV}, on the left, we represent some typical examples of forced solutions which vanish on the right of the obstacle $\delta V$. In these numerical simulations, we work with the exact solutions of the GPE (not limited to linear order) and with well localized obstacles given by 
\begin{equation}
\delta V(z) = \frac{\bar \delta_V}{ \cosh^2[\kappa_p(z-z_p)]}.
\label{dV} 
\end{equation}
For $\bar \delta_V \ll V_0(z = 0)= 3/4 + \kappa^2/9$ (when $\bar{\rho}= \rho_c$ in our unit system), in agreement with the linear treatment of appendix~\ref{App:smallobs}, we observe that the undulation has a sinusoidal shape of wave number $k_0$, and a phase fixed by $z_p$ which is independent of the amplitude. When reaching the nonlinear regime, the wave number decreases with the amplitude, as given by the Weierstrass function of Eq.~\eqref{eq:weierstrass}. Notice that the non-linear regime is reached when $A_f \sim d / \rho_p$. For higher values of $A_f$, pairs of sonic horizons appear due to the undulation, as in the dotted curve in the figure. 

We now consider the general solutions of Eq.~\eqref{eq:forcelineqrho} which are asymptotically bounded on both sides. To satisfy this condition, only the exponentially decaying solution must be kept in the subsonic side. As a result, the space of bounded solutions is of dimension one, and fully characterized by the net amplitude $A_r$ of the oscillations in the asymptotic supersonic region. To define $A_r$ unambiguously, we introduce the amplitude $A_h$ of the oscillations of the {\it free} solution: 
\begin{equation}
\label{eq:corleymode}
\begin{split}
\delta \rho_{\rm hom} \underset{\rm supersonic}\sim A_h \sin [k_0 z + \varphi_h] .
\end{split}
\end{equation}
The phase $\varphi_h$ is fixed modulo $\pi$ 
by the requirement that $\delta \rho_{\rm hom}$ is bounded in the subsonic side. In a linear treatment, the sign of $A_h$ is a true $\mathbb{Z}_2$ symmetry~\cite{Coutant:2012mf}. This degeneracy is lifted when considering the Weierstrass function of Eq.~\eqref{eq:weierstrass}: with our convention, $A_h>0$ corresponds to the linearized limit of the shadow soliton solution, whereas $A_h < 0$ describes the soliton solution. As was found when studying the black hole laser instability~\cite{Michel:2013wpa}, we shall see that the former describes the most stable solution.~\footnote{The recent results obtained by J.~Steinhauer indicate that the $\mathbb{Z}_2$ symmetry is indeed broken at late time, as the mean density displays an oscillatory pattern, see Fig.~2 in~\cite{BHLaser-Jeff}. It would be interesting to verify if the sign of these oscillations corresponds to that of the shadow soliton.}

Combining the free solution with Eq.~\eqref{eq:deltarhoforced}, on the asymptotic supersonic right side, the resulting undulation is 
\begin{equation}
\begin{split}
\delta \rho_r \underset{\rm supersonic}\sim A_r \sin [k_0 z + \varphi_r],
\end{split}
\end{equation}
where 
\begin{equation}
\begin{split}
A_r \ep{i \varphi_r} = A_h \ep{ i \varphi_h}+ A_f \ep{ i \varphi_f}. 
\end{split}
\end{equation}
On the right panel of Fig.~\ref{fig:rhoforvarV}, we represent different profiles of the density $\rho(z)$ for three different values of the amplitude $A_h$ of the free undulation added on top of a given forced undulation with a potential of the form Eq.~\eqref{dV} with $\bar \delta_V = 0.03$, $z_p = 0$, $\kappa_p = 5$, and for the reference potential of Eq.~\eqref{eq:V} with $\kappa = \bar \rho/\rho_c = 10 d = 1$. For increasing values of $A_h$, we first observe a diminution of the resulting amplitude $A_r$, followed by an increase for $A_h > A_f \cos(\varphi_h - \varphi_f)$. The minimum of $A_r$ is nonzero, except when the phase shift $\varphi_h - \varphi_f = \pi$. We also observe a progressive change of the resulting phase.

\section{Condensation of ultra-low frequency phonons and suppression of infrared instability.}
\label{sec:lin}

\subsection{Infrared instability in analogue white hole flows }

We first recall the basic properties of the scattering of phonon modes in trans-sonic stationary flows, such as those studied in the former section. The essential point is the presence of a parametric amplification of counter-propagating modes which diverges in the zero-frequency limit. Generalizing~\cite{Mayoral:2010ck,Coutant:2012mf}, we then explain why this implies that incoming (random) low-frequency phonons produce an undulation, the amplitude of which diverges in this limit. We shall see that this undulation precisely corresponds to the free undulation of Eq.~\eqref{eq:corleymode}. The infrared divergence reveals that analogue white hole flows are unstable against ultra-low frequency perturbations, even though they are dynamically stable, as is explained in appendix~\ref{app:dynstab}.

\subsubsection{Scattering of low frequency phonon modes}

In appendix~\ref{App:homo}, the interested reader will find a review of the key features of phonon modes in homogeneous sub and supersonic flows. Here we consider the mode mixing in inhomogeneous flows, focusing on the zero-frequency limit. More details can be found in Ref.~\cite{Macher:2009nz}. Because the flow is time-independent, stationary linear modes with different frequencies $\omega = i \partial_t $ do not mix among each others when scattered on the inhomogeneous flow. As a result the $S$-matrix can be analyzed at fixed frequency $\omega$ which can be taken positive.

The first important point is to identify the number of independent asymptotically bounded modes (ABM) at fixed $\omega$. Because the background flow crosses the sound speed, for low frequencies, i.e., $0 < \omega < \omega_{\rm max}$ where $\omega_{\rm max}$ is the critical frequency of Eq.~\eqref{eq:ommax}, there are three ABM, and not four as was the case in a homogeneous supersonic flow. The reason is that the two extra modes with negative energy described in appendix~\ref{App:homo} are no longer independent because they are reflected into each other near the sonic horizon as they do not propagate in a subsonic flow. 
 
To describe the mode mixing, one should consider two mode bases: that formed by asymptotic incoming modes that shall be scattered near the sonic horizon, and that formed by outgoing modes that have been scattered. Interestingly, in trans-sonic analogue white hole flows, the three incoming modes are all \enquote{hydrodynamic}, as their wave-vectors $k_\omega^i$ vanish as $\omega \to 0$. One of them, denoted as $\phiin$, is co-propagating in the frame of the fluid, and has a positive norm. As it crosses the sonic horizon region without being significantly affected, it will play a marginal role in the infrared instability we shall describe. The other two are counter-propagating modes, and have opposite norms. The positive-norm one is noted $\phiina$ and the negative-norm one $\phiinb$. (The precise meaning of these notations is recalled in appendix~\ref{App:homo}.) The latter describes phonon excitations with negative energy $ - \hbar \omega$. For low frequency, both of them are strongly amplified when propagating across the sonic horizon region, and are responsible for the infrared instability. 

When considering the three outgoing modes, one recovers the hydrodynamic, co-propagating, positive norm mode that we denote by $\phiout$. The two others have opposite norms, and are now {\it dispersive} modes as their wave-vectors remain finite and opposite to each other in the limit $\omega \rightarrow 0$. Moreover this wave vector is exactly $\pm k_0$ of Eq.~\eqref{eq:defku}. We denote as $\phiouta$ the positive-norm one, and $\phioutb$ the negative-norm one. 

The linearity of Eq.~\eqref{eq:BdG} and the conserved scalar product of Eq.~\eqref{eq:Krein} imply that the two sets are related by 
\begin{equation}
\begin{split}
\label{eq:scat} 
\begin{pmatrix}
\phiina \\
\phiinb \\
\phiin 
\end{pmatrix} 
= \begin{pmatrix}
\Saa & \Sab & \Sac \\
\Sba & \Sbb & \Sbc \\
\Sca & \Scb & \Scc
\end{pmatrix} 
\cdot \begin{pmatrix}
\phiouta \\
\phioutb \\
\phiout 
\end{pmatrix} 
\end{split}
\end{equation}
where the $3 \times 3$ S-matrix is an element of $U(1,2)$ since the negative energy mode has a negative norm. The S-matrix thus describes an anomalous scattering, as can be seen from the unitarity relation associated with the second line: 
\begin{equation}
\begin{split}
\label{eq:scat2}
 |\Sbb|^2- |\Sba|^2 - |\Sbc|^2= 1.
\end{split}
\end{equation}
As explained in~\cite{Macher:2009nz}, $\abs{\Sba}^2$ ($\abs{\Sbc}^2$) describes the pair creation of two counter- (a co and a counter) propagating phonons. When the initial phonon state is vacuum, $\abs{\Sba}^2$ ($\abs{\Sbc}^2$) thus gives the number of counter- \mbox{(co-)} propagating positive energy phonons spontaneously emitted near the sonic horizon. When the background flow is smooth, i.e., $\kappa \xi/c \ll 1$, monotonic and contains no undulation, to great precision~\cite{Macher:2009nz,Finazzi:2010yq,Finazzi:2012iu}, $n_{-\omega} = \abs{\Sba}^2 $ follows a Planck law, with a temperature equal to $T = \kappa/2\pi$, which is the analogue of Hawking's prediction~\cite{Unruh:1980cg}. When the inequality $\kappa \xi/c \ll 1$ is no longer satisfied, the temperature is no longer given by this simple expression, but the spectrum remains Planckian in that for low frequencies, $\abs{ \Sbb }^2 $ and $ \abs{ \Sba }^2$ diverge as 
\begin{equation}
\begin{split}
\label{eq:scat3}
\left\lvert \Sbb \right\rvert^2 \sim \left\lvert \Sba \right\rvert^2 \sim T_0/\omega \gg 1,
\end{split}
\end{equation}
while $\left\lvert \Sbc \right\rvert^2 $ remains finite. 

It is of value to complete this analysis by considering the behavior of the nine coefficients of the S-matrix in the zero-frequency limit. Because of unitarity relations, the S-matrix only contains four independent real quantities, as discussed in~\cite{Busch:2014bza}. This restriction is particularly efficient in the zero-frequency limit where only six out of nine coefficients diverge as $1/\sqrt{\omega}$. In fact, for $\omega \to 0$, the leading coefficients are 
\begin{equation}
\begin{split}
\label{eq:scat4} 
\begin{pmatrix}
\phiina \\
\phiinb \\
\phiin 
\end{pmatrix} 
\sim \sqrt{\frac{T_0 }{ \omega}} 
\begin{pmatrix}
X & X & 0 \\
1 & 1 & 0 \\
\delta X & \delta X & 0
\end{pmatrix} 
 \begin{pmatrix}
\phiouta \\
\phioutb \\
\phiout 
\end{pmatrix} .
\end{split}
\end{equation}
One first notices that by a choice of the arbitrary phases of in and out modes, these coefficients can be taken real and positive. Secondly, unitarity implies $ X^2 + (\delta X)^2 = 1$. In addition, one has generally $\delta X \ll 1$, which means that the co-propagating mode plays a subdominant role in the scattering. In fact in the absence of coupling between co and counter propagating modes, one has $X = 1$. Hence $1- X^2 \neq 0$ accounts for the (weak) coupling with the co-propagating mode.

As a result, to characterize the deviations of the spectrum due to undulations, and the suppression of the IR divergence, it is both convenient and sufficient to study the effective temperature $T_\omega$ defined by
\begin{equation}
\begin{split}
\label{eq:Teff}
{\abs{\Sba}^2} = \frac{1}{e^{\omega/T_\omega} - 1}.
\end{split}
\end{equation}

\subsubsection{Undulations and condensation of low frequency phonons}

Equation~\eqref{eq:scat4} implies that for $\omega \to 0$, up to an overall factor, the three incoming modes possess the same limit
\begin{equation}
\begin{split}
\label{eq:undu1}
\phi_{\omega \to 0}^{\rm in}(z) \sim \left (\frac{T_0}{\omega}\right )^{1/2} \Phi(z) , 
\end{split}
\end{equation}
where $\Phi(z) = \phi_{\omega \to 0}^{\rm out}(z) + \varphi_{\omega \to 0}^{\rm out}(z)$. Notice that the relative phase between $\phi_{\omega \to 0}^{\rm out}$ and $\varphi_{\omega \to 0}^{\rm out}$ (which vanishes in the mode basis we adopted) is fixed by the condition that the modes are asymptotically bounded. Hence $\Phi(x)$ identically corresponds to the free undulation of Eq.~\eqref{eq:corleymode} 
(including its sign ambiguity).

Equation~\eqref{eq:scat4} also implies that (possibly random) incoming modes, or even quantum fluctuations, produce a formally infinite number of low frequency phonons with wave number $\pm k_0$. As explained in appendix~\ref{App:homo}, the group velocity $v_{\rm gr} = 1/\partial_\omega k_\omega$ of these phonons is negative, so they propagate from the sonic horizon in the supersonic (left) region. 

With more precision, assuming that the initial phonon state is characterized by the distributions $n_\omega^{\rm in}$, $n_{-\omega}^{\rm in}$, $n_\omega^{\rm in, co.}$ of the three types of phonons, the final number of low frequency counter propagating phonons (of positive and negative energy) is 
\begin{equation} 
\begin{split}
N_{\rm low} &\sim \int_{\omega > \epsilon} d\omega \frac{2T_0}{\omega} \left(n_{-\omega}^{\rm in} +1 + X^2 n_\omega^{\rm in} + \delta X^2 n_\omega^{\rm in,co} \right) ,
\end{split} 
\end{equation}
where $\epsilon > 0$ is an IR cut-off. To give a physical meaning to this expression, one should consider the number of phonons emitted during a lapse of time $\Delta t$ after the formation of the sonic horizon. In that case, one gets $\epsilon \sim 1/\Delta t$. Depending on the low frequency behavior of the initial distributions, $N$ increases either as $\ln \Delta t$ when $n^{\rm in} \sim \text{constant}$, or as $\Delta t$ when $n^{\rm in}\sim 1/\omega$ as in a thermal state~\cite{Mayoral:2010ck}. 

In any case, when working with a trans-sonic white hole background flow, the linear treatment based on the Bogobuliov de Gennes equation implies that an infinite number of low frequency phonons would be emitted. Strictly speaking, this does not make sense. But, as in Ref.~\cite{Pitaevskii1984} for the case of rotons, it implies that these phonons will condense, and form an undulation with a profile given by Eq.~\eqref{eq:undu1} which will progressively affect the background flow since its amplitude grows. Therefore to determine the late time evolution, it is necessary to compute the modifications of the scattering coefficients of Eq.~\eqref{eq:scat} when taking into account the presence of this undulation. (Whereas Ref.~\cite{Pitaevskii1984} computed the modified spectrum on top of the layered structure, we here compute the modified S-matrix on top of the undulation. The reason for this difference is that the instability we are dealing with is not so much due to the existence of negative-energy modes, but to the infinite amplification of low-frequency modes scattered on a sonic horizon.) 

\subsection{Suppression of the infrared instability}

We now consider the central issue of this paper. It concerns the effects of undulations on the low frequency behavior of the coefficient of Eq.~\eqref{eq:Teff}. The relevant parameters that characterize an undulation are its amplitude $A_r$, its phase $\varphi_r$, and its length. A priori, this length is not an adjustable parameter as it should be taken to be infinite (if one works in truly stationary settings). However we shall see that the limit of long undulations is singular. Therefore, it is appropriate to introduce a profile $D(z)$ that controls its extension and the way its amplitude is damped. This is also justified from the point of view of possible experiments, where the undulation is usually damped because of dissipation. Although our description does not explicitly take dissipative effects into account in the wave equation, using a finite length should correctly describe the main consequences. In what follows we shall work with 
\begin{equation}
\begin{split}
D(z) = \frac{\tanh[\sigma(z-z_{\rm end})]+1}{2} .
\end{split}
\end{equation}
We shall see that the behavior of the low-frequency temperature in the limit $z_{\rm end} \rightarrow \infty$ depends on $\sigma$. Hence the limit of an infinite undulation is not devoid of ambiguity and should be handled with care. The number of oscillations $N$ in the supersonic region, and the number of damped oscillations $\na$ are fixed by
\begin{equation}
\begin{split}
z_{\rm end} = - 2\pi N / k_{0},\quad \sigma = k_{0} / \pi \na, 
\label{lenU}
\end{split}
\end{equation}
where $k_{0}$ is given in Eq.~\eqref{eq:defku}. Notice that on the subsonic side, the undulation profile is automatically regular since only the decaying mode is considered. 

Because of the damping factor, flows with undulations are still homogeneous in both asymptotic regions. As a result, the scattering analysis of the former subsection still applies. We are not aware of any analytical method to study the S-matrix for these flows.\footnote{
We tried to analytically evaluate the low frequency limit of the coefficients of Eq.~\eqref{eq:scat} in the steep regime~\cite{Recati:2009ya,Finazzi:2012iu}, to first order in $A_h \ll 1$, and with $N = \infty$. Assuming that modes are regular in the $\omega \to 0$ limit, we showed that the IR instability is suppressed as the Bogoliubov coefficient $\beta_\omega \sim \sqrt{\omega} / A_h$. Unfortunately, when computing the modes with a large value of $N$, we found that, due to resonant effects, modes contain terms in $A_h / \omega$ and $A_h^2 / \omega^2$ which invalidate the assumption of regularity. This indicates that a nonperturbative treatment in $A_h$ is necessary to control the IR physics. We believe this is beyond the scope of the present work. We are currently trying to carry out such a treatment using Bloch waves \cite{kittel2004introduction}.} We shall thus use numerical methods. We first consider the regime where the relative change of the effective temperature is small. As a second step, we study saturation effects, and show  how the infra-red instability is suppressed in the limit where both $N$ and $\na$ are sent to infinity. The numerical methods we use are directly borrowed from~\cite{Macher:2009nz,Finazzi:2010yq,Finazzi:2012iu,Michel:2014zsa}, to which we refer for a precise description. 

\subsubsection{First deviations}

\begin{figure}[t]
\includegraphics[width=\linewidth]{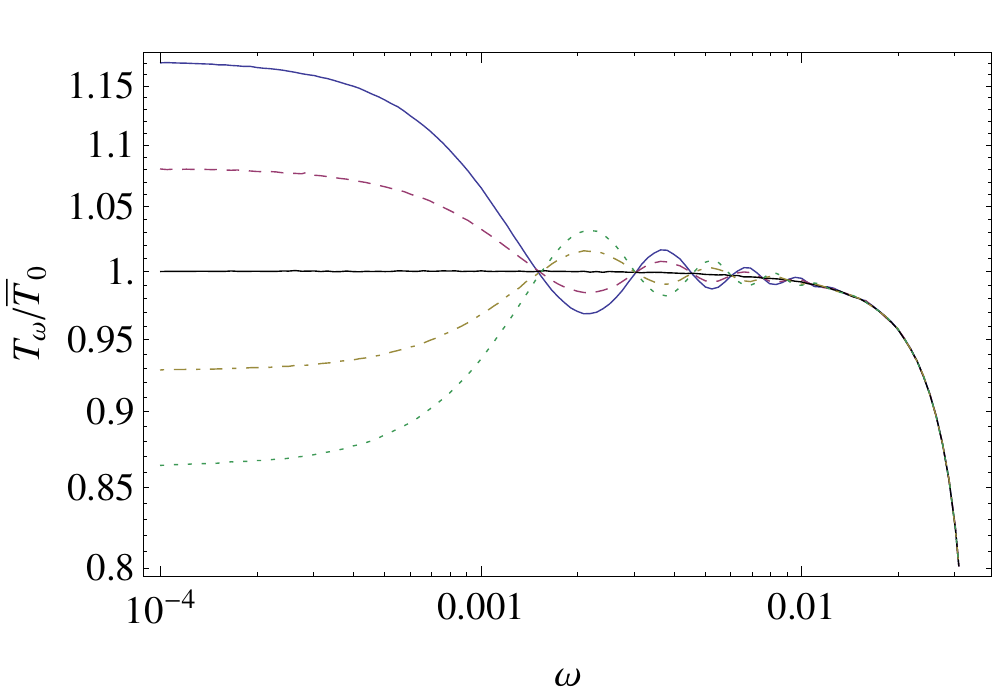}
\caption{
Relative temperature $T_\omega/\bar T_0$ as a function of $\omega$ in log-log scale for flows with {\it free} undulations of different amplitudes on top of Eq.~\eqref{eq:rhotanh} with $\bar \rho/\rho_c = \kappa/3 = 10 d= 1$. The black line represents the temperature in the absence of undulation, and its zero frequency limit defines $\bar T_0$ which sets the vertical scale. The two reduced temperatures are obtained when the free undulation corresponds to a \enquote{shadow soliton}, see appendix~\ref{App:homo}. Their amplitudes are $10^4 A_h = 1$ (dashed-dotted) and $2$ (dotted). The two larger temperatures are obtained for the same amplitudes but the opposite sign. In this regime of weak deviations associated with \enquote{short} undulations [of length here fixed by $n_d = 3$ and $N = 20$, see Eq.~\eqref{lenU}], the temperature change is linear in the undulation amplitude. }
\label{fig:spectra1}
\end{figure}

We first analyze the effect of the amplitude $A_h$ of a free undulation. To this end, in Fig.~\ref{fig:spectra1}, we represent the effective temperature for the reference monotonic flow of Eq.~\eqref{eq:rhotanh}, and with four relatively short undulations, i.e., $N = 20$ and $\na = 3$, of different amplitudes added on top of it. When considering the temperature in the reference flow, we observe that $T_\omega$ of Eq.~\eqref{eq:Teff} is remarkably constant over a large domain of frequencies. This implies that the spectrum is, to a very good approximation, Planckian up to $\omega \lesssim \omega_{\rm max}$ of Eq.~\eqref{eq:ommax} where the power falls sharply. For more details about the spectrum in the absence of undulations we refer to Refs.~\cite{Macher:2009nz,Finazzi:2010yq,Finazzi:2012iu}. 

When considering the four spectra in the presence of undulations, we clearly see that, for low frequency, the relative change of the temperature is linear in $A_h$. We also see that the temperature increases when considering the sign that corresponds to the (linearized version of the) soliton solution, see the description after Eq.~\eqref{eq:weierstrass}, while it decreases when considering the undulation which corresponds to the shadow solution. Interestingly, this is in agreement with what was found in Ref.~\cite{Michel:2013wpa}. In that paper, it was shown that these two solutions govern the stationary backgrounds which characterize the non-trivial density profiles associated with the so-called black hole laser instability~\cite{Corley:1998rk,Leonhardt:2008js}. This dynamical instability occurs in trans-sonic flows with {\it two} sonic horizons separated by a distance $L$. Unlike the present case, the spectrum is characterized by a discrete set of complex frequency modes~\cite{Coutant:2009cu,Finazzi:2010nc}. One can verify that their density is proportional to $L$, while the imaginary part of the frequencies decreases as $1/L$. We therefore conjecture that the present infrared instability should be conceived as the limit where the inter-horizon distance $L$ is sent to infinity. 

\begin{figure}[t] 
\includegraphics[width=1\linewidth]{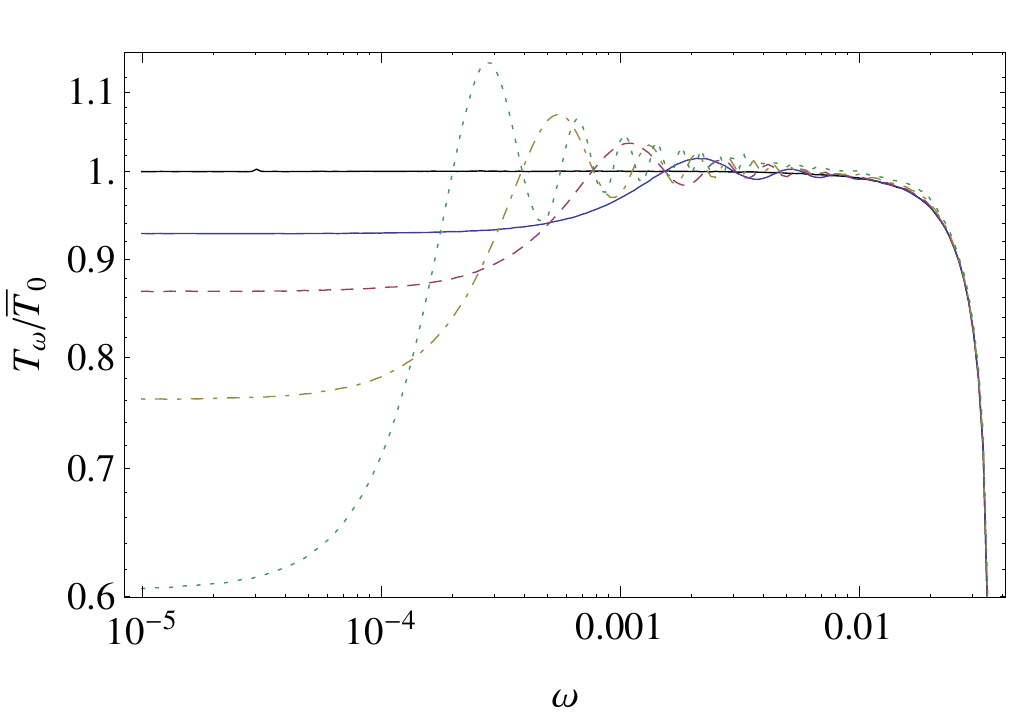}
\caption{As in Fig.~\ref{fig:spectra1}, we represent the relative temperature as a function of $\omega$ in log-log scale for various {\it free} undulations characterized by $A_h = 10^{-4} $, but with different lengths fixed by $\na = 3$, and $N = 20$ (solid), $40$ (dashed), $80$ (dashed dotted), and $160$ (dotted). For large $N$, we notice that the relative change of the temperature becomes much larger than that of the density which is here $\sim 0.01 \%$. We also notice that the period of the oscillations decreases as $1/N$, while their maxima belong to a single envelope. Notice finally that the undulations here considered all describe shadow soliton solutions. }
\label{fig:spectra}
\end{figure}

\begin{figure}[htb]
\includegraphics[width=1\linewidth]{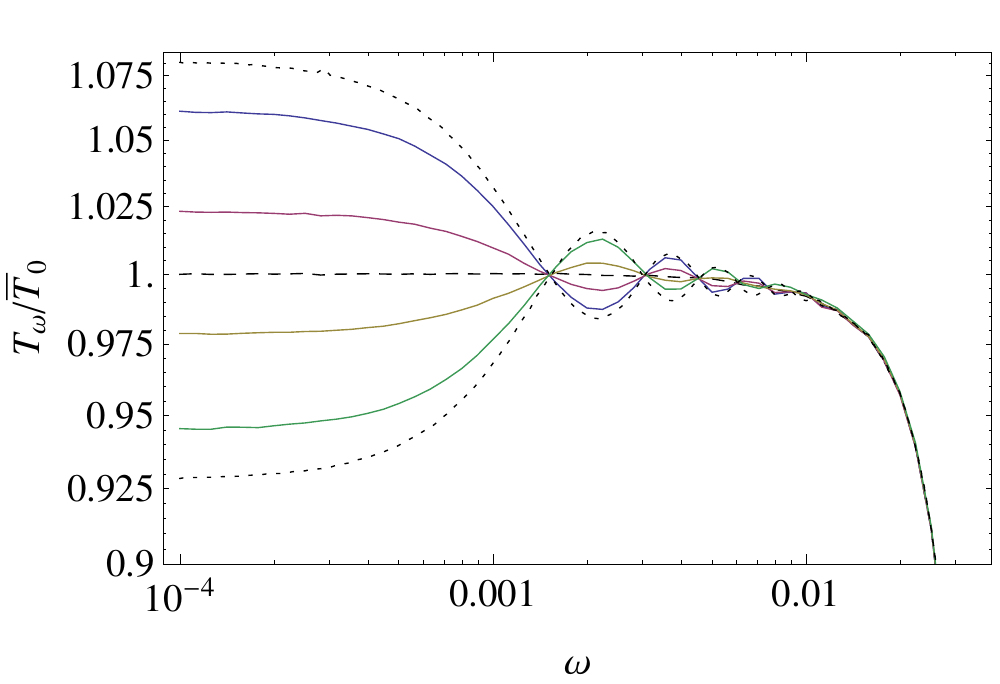}
\caption{In solid lines, we represent the relative temperature associated with four forced undulations obtained by varying the position of the obstacle. For low frequencies, from the top down, the positions are $z_p = 0$ (blue), $ - 1$ (purple), $- 2$ (yellow), $-3$ (green). 
The value of $\bar \delta_V$ is fine tuned so that the amplitude of the undulation is $A_f = 10^{-4}$. The common parameters of the forced undulations are $\kappa_p = 1$, $N= 20$, and $\na=3$. Those of the reference flow are $\kappa/3 = 10 d= 1$. The two dotted lines give the relative temperature for the two signs of the free undulation with the same amplitude and length parameters. One clearly sees that the effective temperature due to forced undulations interpolates between the cases determined by the two signs of the free undulation.}
\label{fig:forced} 
\end{figure}

\begin{figure*}
\begin{minipage}{0.47\linewidth}
\includegraphics[width=\linewidth]{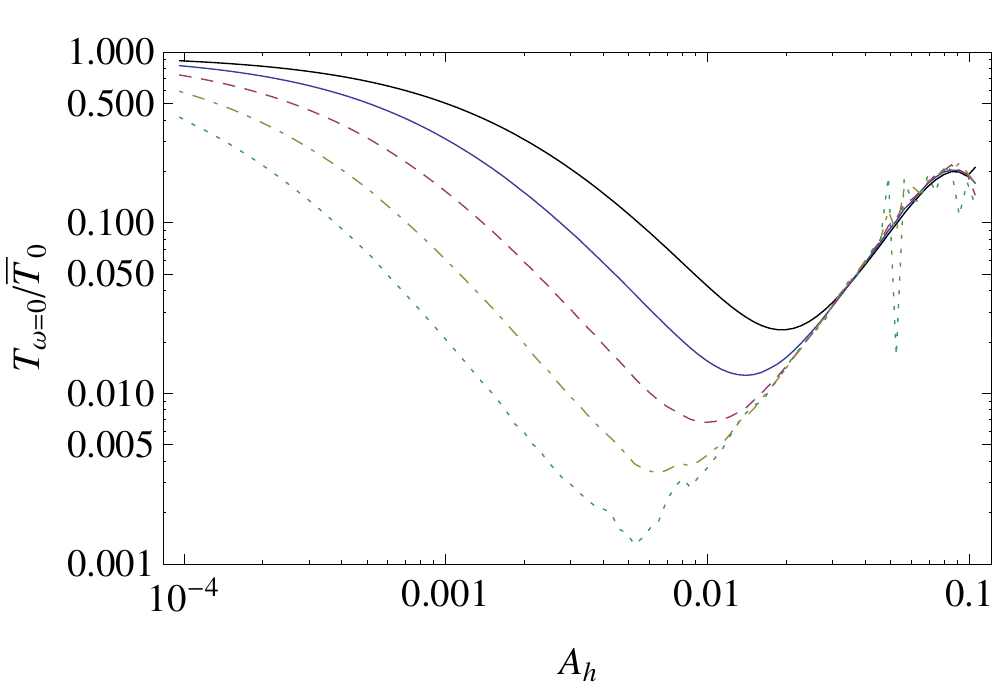}
\end{minipage}
\hspace{0.03\linewidth}
\begin{minipage}{0.47\linewidth}
\includegraphics[width=\linewidth]{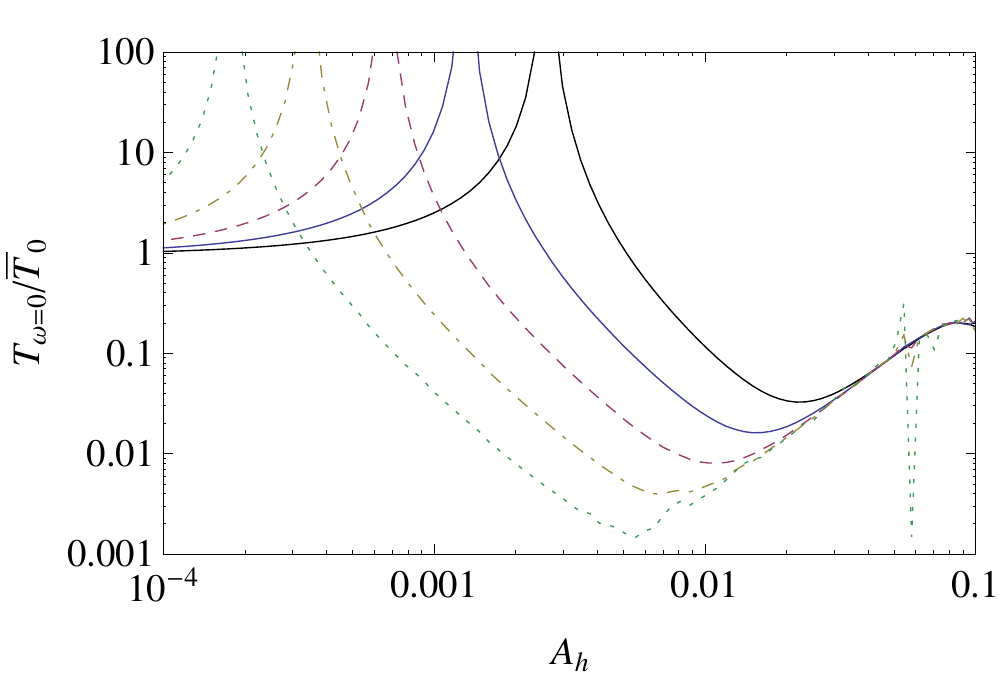}
\end{minipage}
\caption{We represent $T_{\omega = 0}/\bar T_0$, the relative value of the low frequency temperature, as a function of the amplitude of the free undulation $A_h$, for a fixed value of $\na = 3$ and for four values of $N$, namely $20$ (black solid line), $40$ (blue solid line), $80$ (purple dashed line), $160$ (yellow dotted dashed line) and $320$ (green dotted line). The parameters of the background flow are $\kappa/3 = 10 d = 1$, and fix the reference temperature $\bar T_0$. On the left panel, we work with the shadow soliton, which induces a reduction of the temperature
for all vlues of the undulation amplitudes.
 On the right one; we work with the soliton solution which leads to a temperature that diverges for a finite amplitude. As shown in appendix~\ref{app:dynstab}, this indicates that these flows are dynamically unstable for larger amplitudes.  
For large amplitudes both solutions lead to a decrease of the effective temperature. Before reaching the minimum value of $T_{\omega = 0}$, one observes a reduction of the temperature given by Eq.~\eqref{newT}. Instead for larger amplitudes, one sees that the curves become independent of $N$ and grow as $A_h^2$. }
\label{fig:T0A}
\end{figure*}

\begin{figure} 
\includegraphics[width=\linewidth]{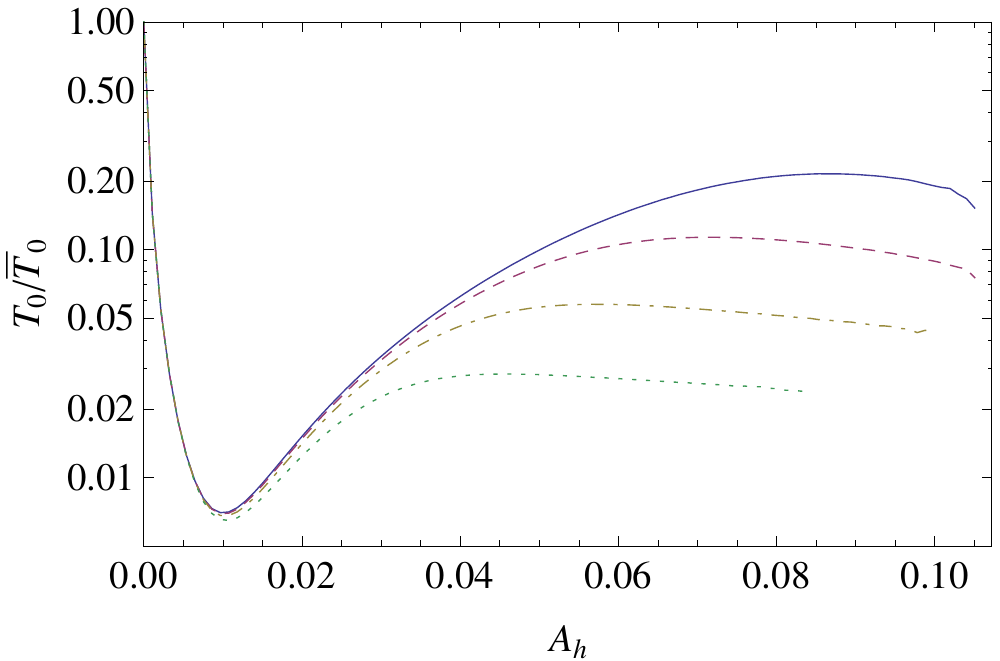}
\caption{We represent the low frequency effective temperature $T_{\omega = 0}$ as a function of the amplitude of a "shadow" undulation for four values of $\na$: $3$ (plain), $6$ (dashed), $12$ (dot-dashed), and $24$ (dotted). The fixed parameters are $N= 80$, $d=0.1$, and $\kappa=3$. We clearly see that the value at which $T_{\omega = 0}$ saturates for $N \to \infty$ decreases as $1/\na$ for $\na \gg 1$. 
}\label{fig:Tna}
\end{figure}

In Fig.~\ref{fig:spectra}, we study the effects of the length of the undulation which is governed by $N$ of Eq.~\eqref{lenU}. We work with a fixed, and small amplitude $A_h = 10^{-4}$, $\na = 3$, and with four values of $N$, namely $20$, $40$, $80$, and $160$. We discover that the relative change of the temperature increases linearly with $N$ and can therefore be much larger than the relative change of the density $\delta \rho$ due to the presence of the undulation. We also observe that the range of low frequency for which the effective temperature is approximately constant decreases as $1/N$. These two observations imply that the limit $N \to \infty$ is singular. 

The linearity of the change of $T_{0} \equiv T_{ \omega = 0}$ in both $A_h$ and $N$ when $A_h {N}$ is not too large implies that in this regime $T_{ 0}$ only depends on their product. In fact a more detailed analysis reveals that, for both signs of $A_h$, 
\begin{equation}
\begin{split}
\label{linsup}
T_{ 0} \sim \bar T_0 \left( 1- \gamma \frac{A_h N}{d} \right), 
\end{split}
\end{equation} 
where $ \bar T_0 $ is the low frequency limit of the temperature when there is no undulation. We checked that $\gamma$ is of order $1$, and weakly depends on $N$. We observe that $T_{ 0}$ depends on $A_h/d$, i.e., the undulation amplitude divided by that associated with the appearance of extra pairs of sonic horizons, see the discussion in the paragraph below Eq.~\eqref{dV}. 

Finally, in Fig.~\ref{fig:forced}, we study the effect of the phase of the undulation by considering forced undulations. Without surprise, when working with a fixed amplitude $A_f$ and various phases $\varphi_f$ by displacing the obstacle, we see that the linear change of the temperature interpolates between the two cases of Fig.~\ref{fig:spectra1} represented by the two signs of the free undulation. To get the temperature change one should simply replace $A_h$ in the above equation by $A_f \cos(\varphi_f - \varphi_h) $. We checked that this interpolation is still found for larger values of $A_f$. As a result, in what follows, we shall no longer consider forced undulations, and study only free undulations with amplitudes of both signs.

\subsubsection{Suppression of infrared instability}

Guided by the results of Fig.~\ref{fig:spectra}, in Fig.~\ref{fig:T0A}, we study the behavior of $T_0 $, as a function of the amplitude $A_h$, for a fixed value of the damping parameter $\na$, and for five values of $N$ equal to $20$, $40$, $80$, $160$ and $320$. On the left panel, we work with the sign corresponding to the shadow soliton. For low values of $A_h N$, we find that $T_0$ decreases with both $N$ and $A_h$, in agreement with Eq.~\eqref{linsup}. Then we see that $T_0$ reaches a minimum and increases with $A_h$, sticking to a curve which is independent of $N$. The critical value of $A_h$ where the minimum of $T_0$ is reached is proportional to $N^{-1/2}$, while the value of this minimum is proportional to $N^{-1}$. Alternatively, when working at fixed $A_h$ and increasing $N$, $T_0$ saturates to a value that we call $T_{\rm sat}(A_h)$. This minimum value goes to zero as $A_h^{2}$ when $A_h \rightarrow 0$. So, for sufficiently small amplitudes the infrared instability is strongly suppressed in the limit $N \rightarrow \infty$.

On the right panel of Fig.~\ref{fig:T0A}, we show the results for undulations corresponding to soliton solutions. 
We see that for small values of $A_h N< 0$, $T_0$ increases with $|A_h|$ as was already observed in Fig.~\ref{fig:spectra1}, see also Eq.~(\ref{linsup}). When increasing $|A_h|$ at fixed $N$, $T_0$ diverges at a critical value of $|A_h|$ proportional to $N^{-1}$, and then follows a behavior similar to that found for the shadow soliton. At this point it should be emphasized that the divergence of $T_0$ reveals the onset of a dynamical instability. As explained in appendix~\ref{app:dynstab}, the divergence is due to the fact that a resonant complex frequency mode crosses the real axis and acquires a positive imaginary frequency. We therefore conclude that only long undulations corresponding to shadow soliton solutions are stable.

Extra simulations at large values of $N$ indicate that the saturation temperature $T_{\rm sat}(A_h)$ found for $N \to \infty$ depends on the value of the damping parameter $\na$. Whenever we could observe it, we found a systematic decrease as $1/\na$, as can be seen in Fig.~\ref{fig:Tna} where the behavior of $T_0$ as a function of $A_h$ is represented for four different values of the damping parameter $\na$. The fact that the maximum of $T_{\rm sat}(A_h)$ is proportional to $1/\na$ indicates that damping the undulation with a slope $\propto 1/\na$ induces some non-adiabatic mode mixing~\cite{Massar:1997en}. From these observations, we strongly conjecture that in the limit where {\it both} $N$ and $\na$ go to infinity, $T_{\omega = 0}$ goes to zero, i.e., the infrared instability is completely suppressed. Combined with the above discussed dynamical stability of shadow undulations, this is the main result of this paper.  It is corroborated by extra numerical simulations performed with $\na \propto N$.

\begin{figure} 
\includegraphics[width=1\linewidth]{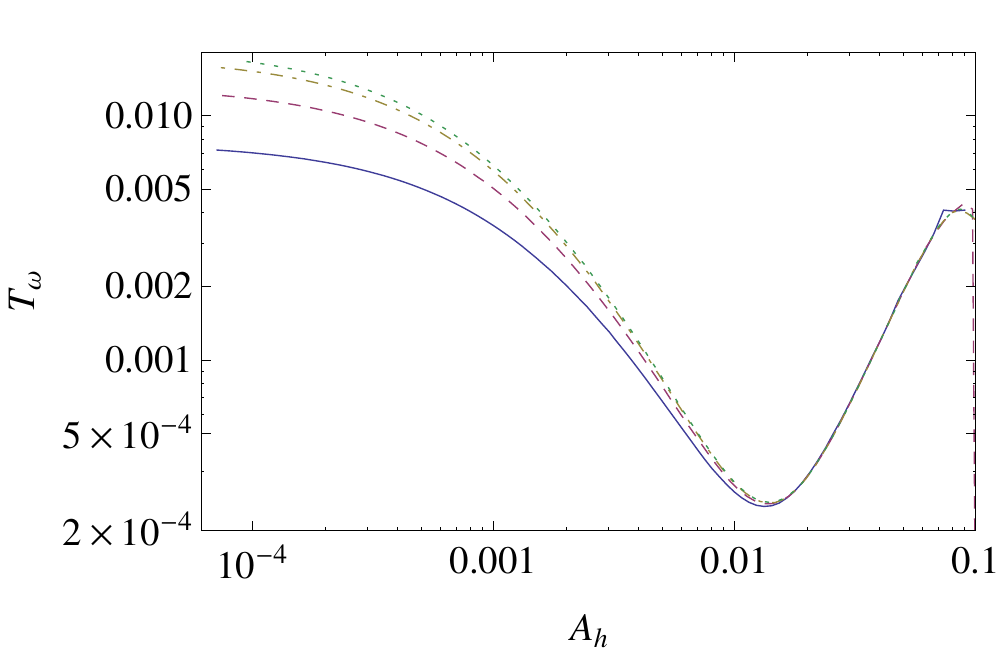}
\caption{We represent the low frequency effective temperature $T_{\omega = 0}$ as a function of the undulation amplitude for $N= 40$, $\na= 3 $, and 4 values of $\kappa$: $1/20$ (blue solid line), $1/10$ (purple dashed line), $1/5$ (yellow dotted dashed line), $0.4$ (green dotted line). For large amplitudes, we observe that $T_{\omega = 0}$ saturates to a value which is independent of the slope $\kappa$. Therefore the surface gravity is not a relevant parameter for the saturation. } 
\label{fig:TofAforkappa}
\end{figure}

So far we worked with values of $\kappa \geq 1$. This is rather large as it amounts to work in the dispersive regime~\cite{Finazzi:2012iu}. In this regime, to a good approximation, $\bar T_0$, the low frequency temperature without undulation, is independent of $\kappa$, and fixed by the asymptotic values of $v$ and $c$ in the sub and supersonic regions. It can thus be computed in a steep approximation, by matching the modes across the sonic horizon~\cite{Recati:2009ya}. When varying $\kappa$, the above picture is basically unchanged for $\kappa \in \left\lbrace 0.05, 5 \right\rbrace$, as can be seen in Fig.~\ref{fig:TofAforkappa}. Based on this, we conjecture that the suppression of the IR instability by undulations in the infinite length limit remains valid for any value of $\kappa$.

We conclude this analysis by two remarks. We notice that in both panels of Fig.~\ref{fig:T0A}, \textit{before} reaching the saturation, $T_0$ depends on $A_{h}$ and $N$ mostly through their product, with a weak dependence in $A_h/N$. Yet, when $|A_h N | \gtrsim d$, Eq.~\eqref{linsup} is no longer valid. For these large values, to a good approximation, we numerically observe the following behavior 
\begin{equation}\label{newT}
\begin{split}
T_0 \propto \frac{\sqrt{ \bar T_0} d^2}{| A_h| N}.
\end{split}
\end{equation}
The power $1/2$ on $\bar T_0$ in this intermediate regime can be understood as an interpolation between the low amplitude regime proportional to $\bar T_0$ and the high amplitude regime independent of $\bar T_0$, see Fig.~\ref{fig:TofAforkappa}. 

It is also of interest to study the properties of the spectrum for $\omega \neq 0$ in this regime of large deviations. Firstly, to a good approximation, we found that the frequency $\omega_1$ where the first maximum of $T_\omega$ is reached, see Fig~\ref{fig:spectra}, is given by
\begin{equation}
\begin{split}
\omega_1 \sim \frac{\omega_{\rm max}}{2 N},
\end{split}
\end{equation}
and has only a weak dependence in the amplitude $A_h$. Secondly, we found that $T_1$, the temperature $T_\omega$ at $\omega_1$, depends on both the amplitude and the number of oscillations, as
\begin{equation}
T_1 - \bar T_0 \sim \bar T_0 |A_h| N^2.
\end{equation}
These equations clearly establish that the limit $| A_h N |\gg d$ is highly singular, and should be handled with care.

\section{Conclusions}
\label{sec:concl}

To prepare the analysis of the scattering of phonon modes in trans-sonic analogue white hole flows, we first studied the space of stationary solutions of the one dimensional GPE subject to a external potential $V$ which is asymptotically constant on both sides. The constant flux $J = \rho v$ was chosen so that the upstream region is supersonic, while the downstream one is subsonic. We considered solutions which are regular and homogeneous in the asymptotic subsonic side. For a given potential, the general solution is then completely characterized by the amplitude of the density modulation in the supersonic side. In addition, when the amplitude of this undulation is small enough, the net amplitude can be decomposed as the sum of a free and a forced solution. It should also be noted that the phase of the free undulation is fixed up to $\pi$. This ambiguity is lifted when considering the bounded solutions of the GPE: one sign of the undulation amplitude corresponds to a shadow soliton solution attached to the sonic horizon, while the other corresponds to a soliton solution. 

In appendix~\ref{app:dynstab} we showed that these white hole flows are dynamically stable, provided the phase of the undulation is close to that of the shadow soliton, and this even when the undulation amplitude is so large that it produces extra pairs of sonic horizons. We also showed that undulations corresponding to soliton solutions become dynamically unstable when their amplitude is large enough. Then, in Sec.~\ref{sec:lin} we studied  the S-matrix which connects the three stationary phonon modes. We showed that in the absence of undulation the spectrum of counter-propagating phonons diverges as $\bar T_0/\omega$ for $\omega \to 0$, see Fig.~\ref{fig:spectra1}. We also explained why this IR divergence produces, in analogue white flows, a condensation of zero-frequency phonons which grows in time, and modulates the flow in the supersonic region by producing a free undulation in the above classification.

To characterize the back-reaction effects due to this undulation, we studied the S-matrix in flows which contain an undulation of given amplitude, extension, and phase. To identify the role of these three parameters, we first analyzed the linear changes of the temperature. For a small undulation (both in amplitude and in length), we showed that the temperature change is proportional to its amplitude, and to its length. As far as the phase is concerned, we showed that the temperature (decrease) increase is largest for the free undulation which corresponds to the linearized version of the (shadow) soliton solution. We believe this is an important result, as it connects the present IR instability to the dynamical instability found when considering flows which are trans-sonic in a finite domain, and which thus cross twice the sound speed~\cite{Corley:1998rk,Finazzi:2010nc,Michel:2013wpa}. The present case can be conceived as the limiting case where the extension of this domain is sent to infinity. This remark is reinforced by studying effects due to larger undulations.

We concluded the analysis by considering the reduction of the temperature due to undulations of large amplitudes and long extensions. All numerical simulations indicate that the IR instability is completely suppressed {\it for any amplitude} when the undulation is taken to be infinitely long, when its phase is close to that of the shadow soliton, and when the profile which governs its \enquote{switching on} is infinitely smooth so as to induce no non-adiabatic transitions. On the contrary, when considering undulations with a phase close to that of the soliton solution, the temperature diverges for a finite amplitude, beyond which we found the above mentioned dynamical instability. We conjecture that the end point evolution of this instability is an undulation with the shadow soliton phase.

To conclude this work, we wish to mention several extensions that would be worth examining. As the analysis was performed in the context of atomic Bose condensates, a first natural extension would be to adapt it to describe undulations in water flumes. We have already started the analysis and we hope to be able to report on this soon. Another interesting context seems to be the fluids of light~\cite{Carusotto:2012vz} where solitons have been recently observed when the condensate is flowing~\cite{Amo03062011}. We hope that the present work could encourage experimental groups to realize experiments aiming at testing our results.

On a more theoretical level, it would be interesting to see if one could use treatments beyond the mean field approximation which could describe the growth in time of the undulation amplitude while taking into account the reduction of low frequency amplification this implies. We believe that effective hamiltonians including one-loop corrections should be able to describe at least the first back-reaction effects due to the undulation.

In this work, we referred several times to the inspiring analysis~\cite{Pitaevskii1984} of the roton condensation in super fluid $^4$He when the flow velocity is slightly above the Landau velocity. We conjecture that there exist connections between this phenomenon, which lead to the concept of super-solid~\cite{RevModPhys.84.759,2010Natur.464..176B}, and the present case. In a preliminary work~\cite{rapportdestage} we studied the ground state of an {\it inhomogeneous} flowing BEC with a dispersion relation showing the same \enquote{roton-maxon} structure~\cite{Baym}, and which crosses the Landau velocity. We numerically observed that the ground state contains a superposition of two undulations associated with the two wave numbers which are the non-trivial zero-frequency roots of the dispersion relation. Indeed because of the roton minimum, two roots are found instead of one in the present case. This result is rather surprising, and needs to be confirmed, because in the case of a homogeneous flow under the hypotheses of~\cite{Pitaevskii1984} only a single wave number enters in layered structures. 

\section*{acknowledgement}

We thank A. M. Kamchatnov, N. Pavloff, and G. Shlyapnikov for interesting discussions in the early stages of this work. We also thank I.~Carusotto for interesting comments on our manuscript. This work was supported by the French National Research Agency under the Program Investing in the Future Grant No. ANR-11-IDEX-0003-02 associated with the project QEAGE (Quantum Effects in Analogue Gravity Experiments).

\appendix

\section{Homogeneous settings}
\label{App:homo}

\subsection{Gross-Pitaevskii equation and non-linear aspects}

The one dimensional time dependent GPE reads
\begin{equation}
\begin{split}
i \hbar \partial_t \psi = -\frac{\hbar^2}{2 m} \partial_x^2 \psi + V \psi + g \left\lvert \psi \right\rvert^2 \psi.
\label{GPEt}
\end{split}
\end{equation}
For stationary solutions $\partial_t \psi=0$, the real and imaginary parts give Eq.~\eqref{GPE1}, where $\rho \equiv \left\lvert \psi \right\rvert^2$ and $m v \equiv \Im \left( \hbar \partial_x \psi / \psi \right)$. 

To understand the structure of the space of stationary solutions, it is worth considering the homogeneous case where $V$ and $g$ are independent of $z$. (In the body of this work we assume $g$ is always constant, while $V$ depends on $z$.) In the unit system defined before Eq.~\eqref{rhoc}, Eq.~\eqref{GPE1} possesses a first integral given by
\begin{equation}
\label{eq:defI1}
\begin{split}
I_1 = 4 V \frac{\rho}{\rho_c}+\frac{\rho'^2}{\rho \rho_c}-\frac{\rho^2}{\rho_c^2}+\frac{\rho_c}{\rho}.
\end{split}
\end{equation}
Asymptotically bounded solutions exist only when 
\begin{equation}
\label{eq:I1constraint}
\begin{split}
4 I_1^3-16 I_1^2 V_0^2-72 I_1 V_0+256 V_0^3+27 \leq 0.
\end{split}
\end{equation}
This defines a finite interval $I_1 \in \left[ I_{1,p}, I_{1,b} \right]$. Its two extremal values give the two homogeneous solutions of Eq.~\eqref{GPE1}: the subsonic one ($\rho > \rho_c$) for $I_1 = I_{1,b}$ and the supersonic one ($\rho < \rho_c$) for $I_1=I_{1,p}$. 

\begin{figure}
\includegraphics[width=\linewidth]{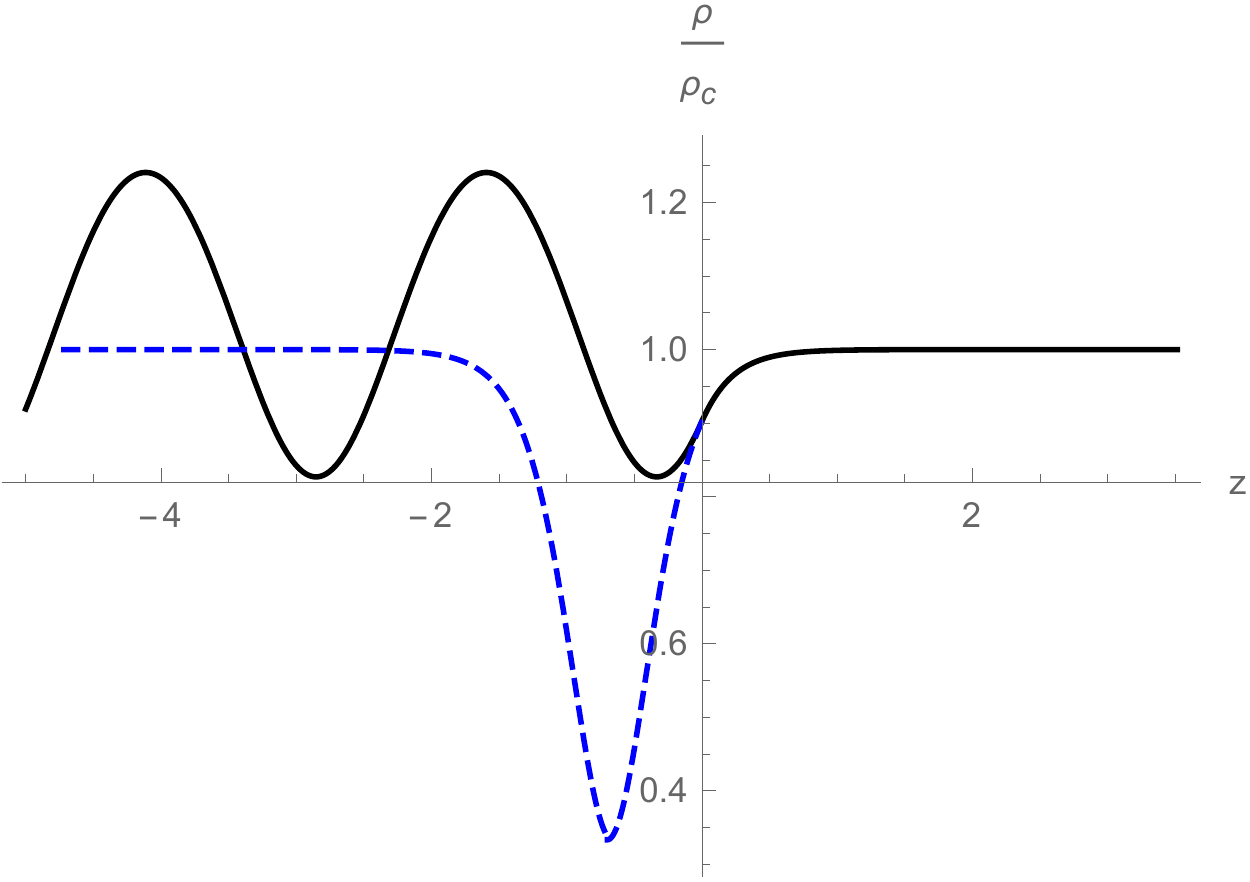}
\includegraphics[width=\linewidth]{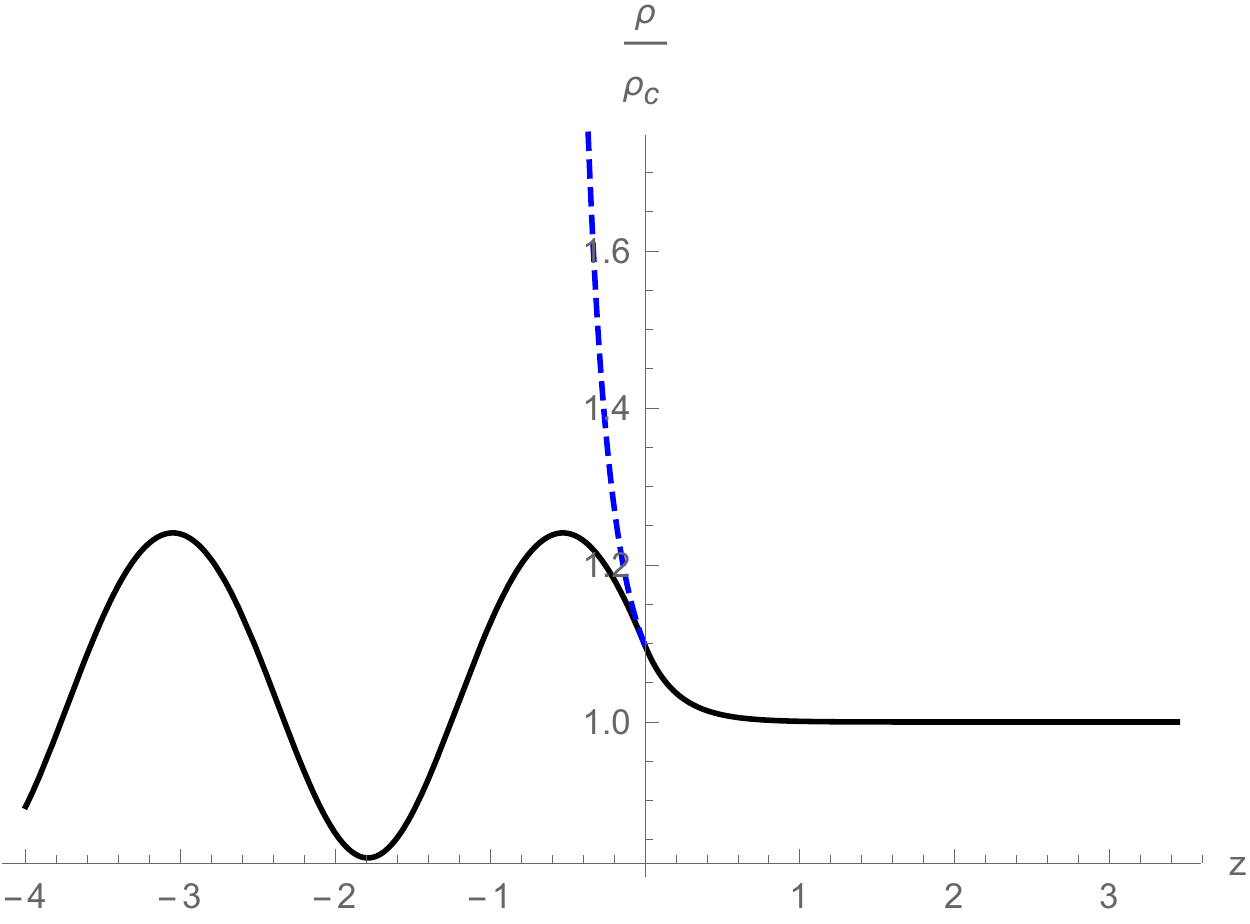}
\caption{Rescaled density profiles $\rho(z)/\rho_c$ of free undulations on the supersonic left region which emerge from a soliton solution (top) and shadow soliton solution (bottom) in case where $V$ and $g$ are piecewise-constant, with a discontinuity at $z=0$. The discontinuity is tuned so that the forced undulation vanishes, and $\rho$ is constant across $z=0$ when there is no free undulation. The parameters are $g(z>0)=8$, $g(z<0)=1$, $\mu(z>0)=28/3$, $\mu(z<0)=7/3$, and $j=\sqrt{8/3}$. In dashed (blue) lines, we represent the full soliton profiles one would obtain without the discontinuity, for parameters equal to their values for $z > 0$.} 
\label{fig:8}
\end{figure}

We now relate bounded solutions of Eq.~\eqref{eq:defI1} to those in the asymptotic homogeneous regions of the potential of Eq.~\eqref{eq:V}. We restrict ourselves to solutions that are bounded on one side, which can then be matched with another one to give globally bounded solutions. The general stationary solution is 
\begin{equation} \label{eq:weierstrass} 
\begin{split}
\rho(z) =& \frac{4\rho_c}{3}V_0 \\
&+ 4 \rho_c \wp \left(z- i c_2;\frac{4 V_0^2}{3}-\frac{I_1}{4},\frac{8 V_0^3}{27}-\frac{I_1 V_0}{12}+\frac{1}{16}\right) ,
\end{split}
\end{equation}
where $c_2$ is half the imaginary period of the Weierstrass function $\wp$, and where $I_1 \in \left[I_{1,p},I_{1,b} \right]$. On the right asymptotic region, we shall only keep the solutions which tend to the homogeneous subsonic one, as only those are asymptotically subsonic. They correspond to $I_1=I_{1,b}$. There are then four branches of solutions determined by the signs of $\rho-\rho_b$ and $\rho'$. When $(\rho - \rho_b) \rho' >0$, the solution grows exponentially at the linear level, thus it shall not be kept. When $(\rho - \rho_b) \rho' <0$, the solution decays exponentially. The solution with $\rho < \rho_b$ is referred to as \enquote{soliton solution}\footnote{In Ref.~\cite{Michel:2013wpa} a distinction is made between two soliton solutions, referred to as \enquote{first} and \enquote{second}. In the present work we use only the first one.} since its continuation towards $z<0$ in a homogeneous condensate contains a soliton, while the solution with $\rho > \rho_b$ is called \enquote{shadow soliton solution}, see Ref.~\cite{Michel:2013wpa} for a recent analysis which details the relations to analogue black hole flows. Each of these branches of solutions has only one continuous free parameter, namely its amplitude. Because of the type of potential we chose in Eq.~\eqref{rhoc}, when this amplitude is small enough to be in the linear regime, these two solutions only differ by a relative minus sign and describe the free undulation of Eq.~\eqref{eq:corleymode}. 
As an illustration, these solutions are represented in Fig.~\ref{fig:8} in the simple case where the potential and two-body coupling are piecewise-constant with a single discontinuity, tuned so that a solution with a homogeneous density exists.

\subsection{Bogoliubov-de~Gennes equation}

\begin{figure}
\includegraphics[width=\linewidth]{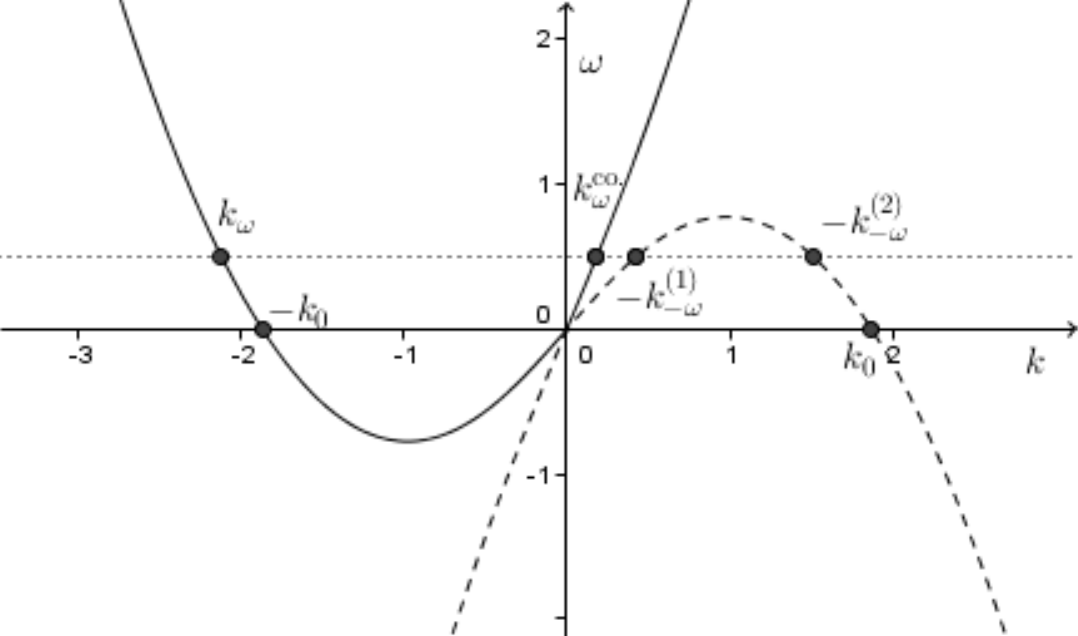}
\caption{Plot of the dispersion relation Eq.~\eqref{eq:disprel} for $\rho / \rho_c = 0.5$, for positive-energy (solid) and negative-energy (dashed) phonons. The wave-vector $k_0$ of the undulation is shown, along with the 4 solutions of the dispersion relation Eq.~\eqref{eq:disprel} for $\omega = 0.5$. The names of these roots are to those of the corresponding modes of Eq.~\eqref{eq:fop2}. One notices that the two roots $-k_{-\omega}^{(1)}, -k_{-\omega}^{(2)}$ live on the branch where the comoving frequency $\Omega = \omega - v k$ is negative. They describe negative energy phonon excitations, with wave numbers $k_{-\omega}^{(1)}, k_{-\omega}^{(2)}$ both negative.
} \label{fig:disprel}
\end{figure}

We work with $\hat \phi$, the relative linear density perturbation operator, defined by $ \hat \Phi = \Phi_{c}(1+\hat \phi)$, where $\hat \Phi$ is the canonical atomic Heisenberg field~\cite{Dalfovo:1999zz}, and $\Phi_c$ is a (possibly inhomogeneous) solution of the GPE which describes the condensed atoms. When working with $\hat \phi$, the Bogoliubov-de~Gennes equation reads~\cite{Macher:2009nz} 
\begin{equation} 
\label{eq:BdG}
\begin{split}
(i \partial_t + i v \partial_x) \hat \phi = mc^2 (\hat \phi + \hat \phi^\dagger) - \frac{1}{2 m \rho} \partial_x \rho \partial_x \hat \phi. 
\end{split}
\end{equation}
Since the background flows we consider are stationary, we can work at fixed frequency $\omega = i \partial_t$. Then, in the unit system defined before Eq.~\eqref{rhoc}, Eq.~\eqref{eq:BdG} reads
\begin{equation}
\label{eq:BdGadim}
\begin{split}
\left ( \omega + \frac{i\rho_c}{\rho} \partial_z \right ) \hat \phi_{\omega}&=\frac{\rho}{2 \rho_c} \left (\hat \phi_{\omega} + (\hat \phi_{-\omega})^\dagger\right )- \frac{1 }{ \rho} \partial_z\rho \partial_z \hat \phi_{\omega}.
\end{split}
\end{equation}
When the flow is homogeneous, the wave vector $k = -i \partial_z$ is also constant, and the dispersion relation reads 
\begin{equation} 
\label{eq:disprel}
\begin{split}
\left(\omega - v k\right)^2 = c^2 k^2 +k^4.
\end{split}
\end{equation}
We remind that the velocity $v$ and sound speed $c$ are related to the density $\rho$ through Eq.~\eqref{eq:vandc}. The dispersion relation is shown in Fig.~\ref{fig:disprel} in the supersonic case $v > c$. To identify the properties of the solutions in the inhomogeneous trans-sonic flows we study, it is appropriate to first consider the set of stationary solutions in homogeneous sub- and supersonic flows. 

When the flow is subsonic, for all $\omega > 0$, Eq.~\eqref{eq:disprel} possesses two real roots: $k_\omega < 0$ and $k_\omega^{\rm co} > 0$. They respectively describe a counter propagating left mover, and a co propagating right mover. For the present analysis, it is convenient to define right and left movers {\it with respect to the flow}, rather than with respect to the lab frame. The two other roots are complex, and respectively describe a growing and a decaying mode. They should be discarded in homogeneous flows. As a result, for $\omega > 0$, the Fourier components $\hat \phi_\omega, \hat \phi_{-\omega}^\dagger$ should be decomposed as~\cite{Macher:2009nz} 
\begin{equation} 
\label{eq:fop}
\begin{split}
\hat \phi_\omega &= \hat a_\omega \phi_\omega + \hat a_\omega^{\rm co} \phi_\omega^{\rm co}, \\
\hat \phi_{-\omega}^\dagger &= \hat a_\omega \varphi_\omega + \hat a_\omega^{\rm co} \varphi_\omega^{\rm co},
\end{split}
\end{equation}
where the operator $\hat a_\omega^i$ ($\hat a_\omega^{i \dagger} $) destroys (creates) a phonon of frequency $\omega$ (a co-propagating one when there is a superscript co, otherwise a counter propagating one). The couples $(\phi_\omega^i, \varphi_\omega^i)$ are normalized using the standard scalar product
\begin{equation}
\begin{split}
\label{eq:Krein}
\left( \phi_\omega^i | \phi_{\omega'}^j \right) &= \int \rho \left[ (\phi_\omega^i)^* \phi_{\omega'}^j - (\varphi_\omega^i)^* \varphi_{\omega'}^j \right ] dx \\
& = \delta_{ij} \delta(\omega - \omega'). 
\end{split}
\end{equation}
For both values of $k_\omega^i$, they are given by
\begin{equation} 
\label{eq:mode}
\begin{split}
\phi_{\omega} &= \frac{u_k}{\sqrt{\rho v_g} }\ep{ i k z}, \\
\varphi_{\omega} &= \frac{v_k}{\sqrt{\rho v_g} }\ep{ i k z} 
\end{split}
\end{equation}
where $v_g = \partial_{k} \omega$ is the group velocity in the lab frame. The factors $u_k$, $v_k$ are 
solutions of
\begin{equation} 
\label{eq:uv}
\begin{split}
u_k^2 - v_k^2=1, \quad \frac{v_k}{u_k} = \frac{2 \sqrt{ \rho \rho_c k^2 + \rho_c^2 k^4 } - 2 \rho_c k^2 - \rho}{ \rho} .
\end{split}
\end{equation}
They depend neither on $\omega$, nor on $v$ because of Galilean invariance. This independence is lost in inhomogeneous flows. 

In a supersonic homogeneous flow, the crucial difference is that there are four linearly independent ABM for $\omega < \omega_{\rm max}$, and two above $\omega_{\rm max}$ where one recovers a situation similar to the subsonic one discussed above. The critical value $\omega_{\rm max}$ is
\begin{equation}
\label{eq:ommax}
\begin{split}
\omega_{\rm max}^2 = {2\frac{\left(\rho_c^3-\rho _p^3\right)^3 \left(\sqrt{8 \rho _p^3 +\rho_c^3} +\rho_c^{3/2}\right)}{\rho _p^4 \rho_c^2 \left(\sqrt{8 \rho _p^3 +\rho_c^3}+3\rho_c^{3/2}\right)^3}}.
\end{split}
\end{equation} 
When $\omega < \omega_{\rm max}$, the two smallest roots are continuously related to the two real ones that existed in subsonic flows. In the limit $\omega\to 0$, the co-propagating root still goes to $0$, whereas the counter propagating root goes to $- k_0$ of Eq.~\eqref{eq:defku}. The two new roots that we call $k_{-\omega}^{(1)}, k_{-\omega}^{(2)}$ both describe phonon excitations which carry a {\it negative} energy $-\hbar \omega$. (This can also be seen from the fact that for $\omega > 0$, these roots are on the negative norm branch of the dispersion relation, for details see Sec.~III.C.2 in \cite{Macher:2009nz}.) This means that supersonic flows are always energetically unstable. We write a subscript $-\omega$ on the corresponding wave-vectors to indicate this fact. Explicitly, in a supersonic homogeneous flow, the expressions of Eq.~\eqref{eq:fop} should be replaced by
\begin{equation} 
\label{eq:fop2}
\begin{split}
\hat \phi_\omega &= \hat a_\omega \phi_\omega + \hat a_\omega^{\rm co} \phi_\omega^{\rm co} + \hat a_{-\omega,1}^{\dagger} (\varphi_{-\omega,1})^*+ \hat a_{-\omega,2}^{\dagger} (\varphi_{-\omega,2})^*, \\
\hat \phi_{-\omega}^\dagger &= \hat a_\omega \varphi_\omega + \hat a_\omega^{\rm co} \varphi_\omega^{\rm co}
+ \hat a_{-\omega,1}^{\dagger} (\phi_{-\omega,1})^* + \hat a_{-\omega,2}^{\dagger} (\phi_{-\omega,2})^* . 
\end{split}
\end{equation}
One also verifies that the group velocity $v_{\rm gr}$ in the lab frame of the two roots which reach $\pm k_0$ for $\omega \to 0$ is oriented upstream. This means that these phonons will propagate away from the sonic horizon in our inhomogeneous flows. 

When considering inhomogeneous flows, the wave vectors $k_\omega^i$ become $z$ dependent. As a result, the various modes mix with each others when working with the exact solutions of Eq.~\eqref{eq:BdGadim}. Moreover, when considering stationary flows which cross once the sound speed, the number of global solutions that are asymptotically bounded on both sides is now three for $\omega < \omega_{\rm max}$. Hence these flows are still energetically unstable. As explained in appendix~\ref{app:dynstab}, the spectrum of ABM remains purely real, which means that these inhomogeneous flows are dynamically stable.

\section{Dynamical stability}
\label{app:dynstab}

In this appendix we study the dynamical stability of trans-sonic flows with and without undulations. We complete our search of complex frequency modes by extending the procedure to quasi-normal modes (QNM), and relate them to the peaks seen in plots of $T_\omega$ versus $\omega$, in a manner similar to that used in~\cite{Finazzi:2010nc,Zapata:2011ze}. The linear dynamical stability of stationary white hole flows (which coincides with that of black hole flows~\cite{Macher:2009nz}) was studied in~\cite{Mayoral:2010ck} in the steep-horizon limit $\kappa \rightarrow \infty$, and in the absence of undulation. Here a dynamical instability is defined as the presence of a solution of the stationary Bogoliubov-de~Gennes equation which obeys outgoing boundary conditions, and which is asymptotically bounded. As we now explain, these modes generally have a complex frequency, with a positive imaginary part. They thus grow exponentially in time.

We define outgoing boundary conditions in the following way. We first choose a convex domain which contains the origin in the complex $\omega$ plane, but none of the other values of $\omega$ for which the dispersion relation Eq.~\eqref{eq:disprel} has a double root in $k$.\footnote{Notice that our definition extends to any domain containing no closed path around any frequency $\omega$ with a negative imaginary part and for which the dispersion relation, seen as an equation on $k$, possesses a double root. One can in principle let this domain include the whole upper complex half-plane, paying attention to possible branch cuts, as when two roots of the dispersion relation cross each other they are either both outgoing or both incoming. In the lower half-plane however, an outgoing root of the dispersion relation may cross an incoming one. This makes our analytical continuation ill-defined if one can make a full turn around a value of $\omega$ where this happens, as an outgoing wave-vector would become incoming after one turn, and conversely. To prevent this, one must restrict to a domain which does not contain any closed loop around such a point. Notice also that the results in the complex lower half-plane in general depend on the choice of the domain. One way to fix the ambiguity is to restrict to star domains with respect to the origin. The results thus obtained are unambiguous, and generalize those found using convex domains. In practice, in our numerical calculations we mostly used domains with a rectangular shape.} In this domain, the roots of Eq.~\eqref{eq:disprel} may be labeled unambiguously so that each root is an analytical function of $\omega$. We then define an outgoing wave as one which, once analytically continued on the real $\omega$-axis, is either outgoing in the usual sense or exponentially decreasing. One can show~\cite{Michel:2013wpa} that in the upper complex half-plane these outgoing modes exactly coincide with the asymptotically bounded modes (ABM), while in the lower half-plane they are unbounded and correspond to QNM. 

Let us now consider a transonic configuration with asymptotically homogeneous regions. We denote as $k_{1,b}$, $k_{2,b}$, $k_{3,b}$, and $k_{4 ,b}$ the wave-vectors in the asymptotic right region, where $k_{1,b}$ and $k_{2,b}$ are outgoing while $k_{3,b}$ and $k_{4,b}$ are not. The wave-vectors in the asymptotic left region are defined similarly, with the index \enquote{b} replaced by a \enquote{p} since we are then in the supersonic region. The outgoing boundary conditions on the right impose to work with linear combinations of the modes of wave-vectors $k_{1,b}$ and $k_{2,b}$. Moreover, asymptotic boundedness on the left imposes that the coefficients before the modes in $k_{3,p}$ and $k_{4,p}$ vanish. Denoting as $W_{i,b}$ (respectively $W_{i,p}$) the two component solution of Eq.~\eqref{eq:BdGadim} which is proportional to $e^{i k_{i,b} x}$ ($e^{i k_{i,p} x}$) at $x \rightarrow \infty$ ($x \rightarrow -\infty$), and expanding them as
\begin{equation}
\begin{split}
W_{i,b} = \sum_j A_{i,j} W_{j,p},
\end{split}
\end{equation}
an outgoing modes exists if and only if
\begin{equation}
\begin{split}
\Delta(\omega) \equiv
\left\lvert
\begin{matrix}
A_{1,3} & A_{1,4} \\
A_{2,3} & A_{2,4}
\end{matrix}
\right\rvert =0.
\end{split}
\end{equation}
We use the following procedure to numerically evaluate the zeros of $\Delta$. Assuming this quantity is a holomorphic function of $\omega$ (except for possible branch cuts and singularities), zeros of $\Delta$ show up through a change of phase by a multiple of $2 \pi$ along a closed path. Assuming there is neither singularity nor branch cut inside the contour, this allows one to count the number of ABM and QNM frequencies it contains. The contour can then be refined to locate these frequencies more precisely, before verifying that $\Delta$ vanishes. 

\begin{figure}
\includegraphics[width=\linewidth]{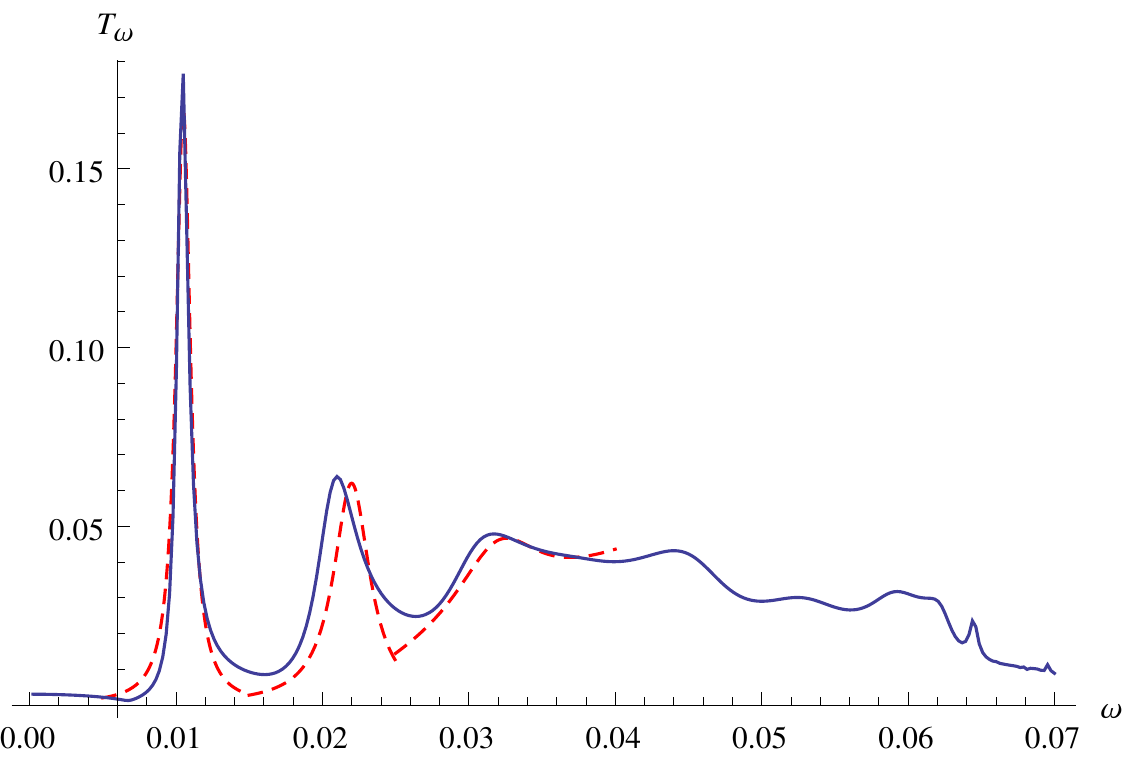}
\caption{Blue, solid: $T_\omega$ as a function of $\omega$ for $d=0.1$, $\kappa = 0.5$, $N=7$, $\na=3$, and $A_r = 0.05$. Red, dashed: Lorentzians with positions and widths fixed by the complex frequencies of the first three QNM for the same parameters, see Eq.~\eqref{eq:dTQNM}. The only free parameters for the two first Lorentzians are their amplitudes, which were fixed to match the height of the corresponding peaks. For the third one a linear function of $\omega$ was added to take the smooth part of $T_\omega$ into account. One can see that Eq.~\eqref{eq:dTQNM} gives a good approximation of the three first peaks. 
}\label{fig:firstQNM}
\end{figure}

We first used this method for a flow of Eq.~\eqref{eq:rhotanh} with $d=0.1$, $\kappa = 0.5$, and with a free undulation (with the sign corresponding to a shadow soliton) added on top with parameters $A_r= 0.05$, $N=7$, and $\na=3$. We found no ABM in the domain we searched, i.e., for $\left\lvert \Re (\omega) \right\rvert < 0.07$ and $\left\lvert \Im (\omega) \right\rvert < 0.05$, indicating that the flow should be (linearly) dynamically stable. As expected given the results of~\cite{Finazzi:2010nc,Zapata:2011ze}, we found a one-to-one correspondence between QNM and the peaks of $T_{\omega}$ visible in Figs.~\ref{fig:spectra1} and~\ref{fig:spectra}. A QNM frequency $\lambda = \omega - i \Gamma$, $\left( \omega_a, \Gamma_a \right) \in \mathbb{R}_+^2$ should give a peak of the form 
\begin{equation} \label{eq:dTQNM}
T_\omega \sim \frac{\delta_a}{(\omega-\omega_a)^2 + \Gamma_a^2},
\end{equation} 
for $\omega \sim \omega_a$. Here $\delta_a$ is a real parameter. Fig.~\ref{fig:firstQNM} shows a comparison between $T_\omega$ and three Lorentzians with positions and widths fixed by the frequencies of the first three QNM. The free parameters in the fit are the three weighs $\delta_a$, three additive constants and linear functions added to take into account the fact that $T_\omega$ does not vanish in the absence of QNM. These additional degrees of freedom hardly change the position of the maxima and the width of the peaks, and thus do not affect our conclusions. The good agreement between these curves tells us two things. Firstly, it indicates that our method to determine the values of $\left( \omega_a, \Gamma_a \right)$ works well and can thus be trusted. Secondly, it also indicates that there are no other complex frequency modes, since these would be seen as unexplained peaks in the function $T_\omega$. Therefore this absence is an other indication for the absence of dynamical instability in the present case.

When considering negative values of $A_r$ which describe undulations close to a soliton solution, we found a similar behavior except for one mode, whose frequency has a vanishing real part. For small values of the amplitude $\left\lvert A_r \right\rvert$, i.e., on the left of the peak shown in Fig.~\ref{fig:T0A}, this mode is a QNM. When reaching the peak of that figure, its frequency crosses the real axis, and for larger amplitudes, this mode describes a dynamical instability. 
Numerical simulations indicate that the imaginary part of the unstable mode goes to a constant in the limit of a long undulations. Therefore, flows with undulations of a large enough negative $A_r$ are dynamically unstable, in a manner similar to the soliton solutions found in background flows experiencing a black hole laser instability~\cite{Michel:2013wpa}. 

When increasing $|A_r|$, the peaks of $T_\omega$ become thinner, and the QNM frequencies move accordingly towards the real axis. We studied values of undulation amplitudes up to $A_r = 0.1$. For this value of the amplitude, the flow possesses extra pairs of black and white horizons in the upstream region, as it is the case in Fig.~\ref{fig:rhoforvarV} left panel for the dotted curve. In spite of these extra horizons, which could have produced some lasing effect~\cite{Corley:1998rk,Coutant:2009cu,Finazzi:2010nc}, we never observed any 
other QNM frequency crossing the real axis. If this occurred, $T_\omega$ would have diverged as $T_\omega \propto (\omega - \omega_0)^{-2}$, something we never saw either for shadow solitons. When decreasing $A_r$, the peaks become broader and the QNM frequencies move away from the real axis. For $A_r = 0$, we found neither ABM, nor QNM. Although not a rigorous proof of linear dynamical stability of undulations with $A_r > 0$, we believe these results are a strong indication for it.~\footnote{In addition, we confirmed them for flows without undulation by solving the time-dependent GPE for $\kappa=0.05$, as was also done in~\cite{Mayoral:2010ck}. We imposed periodic boundary conditions for an integration domain of length $10^3$ and two horizons were separated by a length of $4 \cdot 10^2$ (long enough for the instability coming from the black hole laser effect~\cite{Corley:1998rk,Coutant:2009cu,Finazzi:2010nc,Michel:2013wpa} to be negligible). Adding a random white noise of amplitude $10^{-3}$ to the initial configuration, we still found no exponential amplification after an integration time of $150$. }

\section{Forced zero-mode induced by a small obstacle}
\label{App:smallobs}

In this appendix we compute the quantities $A_f$ and $\varphi_f$ of Eq.~\eqref{eq:deltarhoforced} in the simple cases where the perturbation is located sufficiently far from the horizon, where the flow is homogeneous. For the sake of clarity, we deal separately with the two cases where the obstacle lies in the supersonic or in the subsonic asymptotic regions.
We refer the reader to~\cite{PhysRevLett.97.260403} for the 2-dimensional case.

In the case where the perturbation is located in the supersonic region, Eq.~\eqref{eq:forcelineqrho} reduces to
\begin{equation}
\begin{split}
\label{eq:Aprholin}
\delta \rho''(z)+k_0^2 \delta \rho(z) = 2 \delta V(z).
\end{split}
\end{equation}
In Fourier space, this gives 
\begin{equation}
\begin{split}
\left ( k_0^2 - k^2 \right ) \widetilde{\delta \rho}(k) = 2 \widetilde{\delta V}(k) e^{-i k z_p}, 
\end{split}
\end{equation}
where the Fourier transforms are defined by
\begin{equation}
\begin{split}
\widetilde{\delta \rho}(k) & \equiv \int \frac{dz}{\sqrt{2 \pi}} e^{-i k z} \delta \rho (z), \\
\widetilde{\delta V}(k) & \equiv \int \frac{dz}{\sqrt{2 \pi}} e^{-i k z} \delta V(z+z_p). 
\end{split}
\end{equation}
The shift on $z$ by $z_p$ ensures $\widetilde{\delta V}(k)$ is unchanged when translating the obstacle. The general solution is, imposing $\rho \in \mathbb{R}$: 
\begin{equation}
\begin{split}
\widetilde{\delta \rho}(k) = a \delta (k-k_0) + a^* \delta (k+k_0) + \frac{2 \widetilde{\delta V}(k) e^{-i k z_p}}{k_0^2 - k^2}, 
\end{split}
\end{equation}
$a \in \mathbb{C}$. Taking the inverse Fourier transform gives
\begin{equation}
\begin{split}
\delta \rho(z) = A_h \sin \left(k_0 z + \varphi_h \right) + \sqrt{\frac{2}{\pi }}\int e^{i k (z - z_p)} \frac{\widetilde{\delta V}(k)}{k_0^2-k^2} dk,
\end{split}
\end{equation}
$(A_h, \varphi_h) \in \mathbb{R}$. We recover the free and the forced undulations of Eqs.~(\ref{eq:deltarhoforced},\ref{eq:corleymode}). 

Focusing on the forced solution, we now assume that $\delta V$ is localized and that $\widetilde{\delta V}$ is analytic. In the left asymptotic region, we know from Eq.~\eqref{eq:Aprholin} that the forced undulation is a superposition of plane waves $e^{\pm i k_0 z}$, i.e., only the poles of the integrand contribute. Their contributions are unambiguously defined only after the path around the two poles in the complex plane is chosen. We want the forced zero-mode to vanish on the right of the obstacle where the integration contour must be closed in the upper half-plane to apply the Cauchy theorem. We thus choose a path going slightly above the two poles. For $z \rightarrow -\infty$, we close the integration contour in the lower half-plane, pick the contributions of both poles, and obtain
\begin{equation}
\begin{split}
\int dk &e^{i k z} \frac{\widetilde{\delta V}(k) e^{i k z_p}}{k_0^2-k^2} \\
= & -2 \pi \frac{\abs{\widetilde{\delta V} (k_0)}}{k_0} \sin \left(k_0 (z-z_p) +\arg \left(\widetilde{\delta V} (k_0)\right)\right),
\end{split}
\end{equation}
where we used $\widetilde{\delta V} (-k_0)=\widetilde{\delta V} (k_0)^*$, $\delta V$ being real. We thus obtain $A_f = 2 \sqrt{2 \pi} \abs{\widetilde{\delta V} (k_0)}/k_0$, and $\varphi_f =\arg \left(\widetilde{\delta V} (k_0)\right) - k_0 z_p$. 

Let us now consider the case where the obstacle is put in the homogeneous subsonic region. The forced solution is then given in Fourier space by
\begin{equation}
\begin{split}
\widetilde{\delta \rho}(k) = -\frac{2 \widetilde{\delta V}(k) e^{-i k z_p}}{\lambda_0^2 + k^2} ,
\end{split}
\end{equation}
with $\lambda_0$ real and taken $> 0$. Going back to real space gives
\begin{equation}
\begin{split}
\delta \rho(z) = -\sqrt{\frac{2}{\pi}} \int e^{i k (z-z_p)} \frac{\widetilde{\delta V}(k)}{\lambda_0^2 + k^2} dk.
\end{split}
\end{equation}
The two poles of the integrand are now at $k = \pm i \lambda_0$. Since we want $\delta \rho$ to vanish on the right of the obstacle, we chose the contour which passes above the poles. On the left of the obstacle (but still in subsonic region), one thus has
\begin{equation}
\begin{split}
\delta \rho_f(z) &= -\frac{\sqrt{2 \pi}}{\lambda_0} \\
&\times \left( \widetilde{\delta V}(-i \lambda_0) e^{\lambda_0 (z-z_p)} - \widetilde{\delta V}(i \lambda_0) e^{-\lambda_0 (z-z_p)} \right).
\end{split}
\end{equation} 
This can be matched explicitly to the general solution in the supersonic region, for $z < 0$, in the steep regime limit. Using Eq.~\eqref{eq:forcelineqrho}, a straightforward calculation gives
\begin{equation}
\begin{split}
A_f^2 = -\frac{2 \pi}{\lambda_0^2} \left( \widetilde{\delta V}(-i \lambda_0) e^{-\lambda_0 z_p} - \widetilde{\delta V}(i \lambda_0)e^{\lambda_0 z_p} \right)^2 - \\
 \frac{2 \pi}{k_0^2} \left( \frac{\rho_b}{\rho_p} \right)^2 \left( \widetilde{\delta V}(-i \lambda_0)e^{-\lambda_0 z_p} + \widetilde{\delta V}(i \lambda_0) e^{\lambda_0 z_p} \right)^2, 
\end{split}
\end{equation}
and
\begin{equation}
\begin{split}
\tan (\varphi_u) = \frac{\lambda_0 \rho_p \left( \widetilde{\delta V}(-i \lambda_0) e^{-\lambda_0 z_p}- \widetilde{\delta V}(i \lambda_0) e^{\lambda_0 z_p}\right)}{k_0 \rho_b \left( \widetilde{\delta V}(-i \lambda_0) e^{-\lambda_0 z_p}+ \widetilde{\delta V}(i \lambda_0) e^{\lambda_0 z_p} \right)},
\end{split}
\end{equation}
the sign of $\cos (\varphi_u)$ being that of $-\widetilde{\delta V}(-i \lambda_0) - \widetilde{\delta V}(i \lambda_0)$.

\bibliographystyle{apsrev4-1}
\bibliography{../biblio/bibliopubli}

\end{document}